\documentclass[preprint,12pt]{elsarticle}
\usepackage{amsmath}

\usepackage{amssymb}

\usepackage{lineno}

\newcounter{bla}

\journal{Computer Physics Communications}

\begin{document}

\begin{frontmatter}



\title{CoFFEE: Corrections For Formation Energy and Eigenvalues for charged defect simulations}


\author{Mit H. Naik}
\author{Manish Jain\corref{author}}

\cortext[author] {Corresponding author.\\\textit{E-mail address:} mjain@physics.iisc.ernet.in}
\address{Centre for Condensed Matter Theory, Department of Physics, Indian Institute of Science, Bangalore 560012, India}

\begin{abstract}
Charged point defects in materials are widely studied using Density Functional 
Theory (DFT) packages with
periodic boundary conditions. The formation energy and defect level computed
from these simulations need to be corrected to remove the contributions from the
spurious long-range interaction between the defect and its periodic images.
To this effect, the CoFFEE code implements the Freysoldt-Neugebauer-Van de Walle
(FNV) correction scheme.
The corrections can be applied to charged defects in a complete range of material
shapes and size: bulk, slab (or two-dimensional), wires and nanoribbons. The code is
written in Python and features MPI parallelization and optimizations using the Cython
package for slow steps.
\end{abstract}


\end{frontmatter}



{\bf PROGRAM SUMMARY}

\begin{small}
\noindent
{\em Program Title:}    CoFFEE                                \\
{\em Program obtainable from:}  \verb|http://www.physics.iisc.ernet.in/~mjain/pages/software.html| \\
{\em Journal Reference:}                                      \\
{\em Catalogue identifier:}                                   \\
{\em Licensing provisions:} Open source BSD License                \\
{\em Programming language:}  Python                              \\
{\em Computer:} Any computer with Python installed. The code has been tested with Python2.7 and Python3.6.          \\
{\em Operating system:} Unix/Linux/Windows                    \\
{\em RAM:} 5-1000 MB (dependent on system size)           
{\em Classification:}  7.1, 7.3                             \\
{\em External routines/libraries:} numpy, scipy, mpi4py, matplotlib \\
{\em Nature of problem:} Most electronic structure codes based on Density Functional Theory 
use periodic boundary conditions. This leads to spurious electrostatic interactions during simulation of 
charged defects, which affects the computed defect formation energy and the defect eigenvalue. 
{\em Solution method:} We implement the Freysoldt-Neugebauer-Van de Walle (FNV) correction scheme to correct 
the defect formation energy and eigenvalues. Our implementation can be applied to charged defects in 3D bulk materials
as well as materials having 2D and 1D geometries.\\
{\em Running time:} 1-600 minutes (depends on the number of processors and system size)\\
\end{small}

\section{Introduction}

Point defects, ubiquitous in materials, influence their electrical and optical properties.
First principles electronic structure calculations
have proven to be vital in understanding and predicting the role of defects
\cite{RMP_Walle,PRB_Andrei,PRL_Jain,PRB_Choi,PRB_Bjaalie,PRA_Diallo,PRA_Oba,PRB_Lee,
PRB_Wang,PRB_Sun,APL_Daniel,JAP_Neugebauer,PE_Mota,JCTC_Chen,Manjanath.CPL,PRL.Oif,
PRB.Tiago,PRB.Canning}.
Defect engineering to enhance or suppress certain characteristics of materials 
often rely on such simulations for inputs \cite{PRB_Komsa3,JPCC_Talat,ARMR_Tuller,
AFM_Zhao,JPCC_Nowotny,Singh.ACSNano,PRB.Bhowmick}.

Intrinsic point defects in materials are generally found in low concentrations, about one in 
a million atoms in 3D solids to one in a thousand atoms in 2D materials. Simulations 
thus attempt to study isolated defects in materials. 
However, a large number of first principles calculation codes based on Density Functional 
Theory (DFT) employ periodic boundary conditions. As a result,
in order to simulate and understand isolated defects, super cells are constructed.
The size of the super cell is chosen 
to minimize the overlap of the defect wavefunction with its periodic image in 
neighboring cells \cite{RMP_Walle,PRL_FNV,PRB_KRP,PRB_Oba,PRB_Dabo,JPCSSP_Gillian,PRB_Hine}. 
The computed formation energy of non-shallow neutral defects 
in such calculations is found to converge quickly with the super cell size.

Charged point defects are simulated by artificially introducing a compensating 
uniform background charge to avoid the divergence in the electrostatic energy. 
The formation energy of charged defects show slow 
convergence with super cell size due to the Coulomb interaction between the defect 
charge and its periodic images. The super cell sizes necessary to completely overcome these 
spurious defect-defect interactions are computationally intractable.
The defect energy levels in the gap are also affected similarly,
and shows slow convergence with super cell size 
\cite{PRB_KRP,JPCM_Chen,PRB_CP}. 
Furthermore, since the uniform background charge is artificially introduced, 
an absolute reference for the electrostratic potential is undefined. 
The potential in the defect supercell thus needs to be aligned with the bulk potential, 
in accordance with the chosen reference of the Fermi level with respect to the bulk 
VBM. 

Several \emph{a posteriori} correction schemes have been developed to tackle this 
issue \cite{PRL_FNV,PRB_Oba,PRB_Payne,PRB_Zunger,PRB_Dabo,PRL_Zhang,CPC.Chan}. 
Most of the correction schemes involve solving the Poisson equation for a model 
system, and aligning of potentials, to derive the correction to the formation energy. 
In particular, the scheme proposed by 
Freysoldt, Neugebauer and Van de Walle (FNV) has gained a lot of 
popularity owing to its consistency in deriving 
accurate corrections for charged defects in numerous materials \cite{PRL_FNV}. 
This 
scheme has been extended to low-dimensional systems as well, and shown to 
perform well \cite{PRL_KP,PRB_Noh,PRX_Komsa,PRB_Komsa2,PRB_Park,arxiv.Kax,PRB.Sensoy}. 
However, a generalized correction scheme implementation that works with bulk as well 
as low-dimensional systems is absent in the various DFT
packages, or as an independent package \cite{arxiv.Bro,CPL.PyDEF}. 
The recent independent packages are restricted to compute the corrections for charged 
defects in bulk systems alone \cite{arxiv.Bro,CPL.PyDEF}. 

We present a complete electrostatic corrections package, CoFFEE: Corrections For Formation Energy 
and Eigenvalues for charged defect simulations. The package is applicable to 
charged defects in materials ranging bulk solids, interfaces, surfaces/slabs, 
two-dimensional (2D) materials, nanowires and nanoribbons. These materials can be 
classified according to the number of periodic directions, into 3D (bulk)
, 2D (slabs, 2D materials) and 1D (nanowires, nanoribbons) systems. We implement a 
generalized Poisson solver based on the FNV correction scheme with 
a gaussian model charge distribution. Tools to 
compute the potential alignment terms in the FNV correction scheme are also provided 
with the package. The code is written entirely in Python \cite{Rossum}. 
We use Message Passing Interface (MPI) to parallelize 
the code and Cython \cite{cython} to accelerate slow steps. Our implementation can be 
used alongside any DFT package to obtain an \emph{a posteriori} correction for the 
formation energy and the defect level position 
in the gap, for the charged defect being simulated.  

\section{Theoretical framework}

The formation energy of a neutral defect in a material is given by:
\begin{equation}
 \label{eqn1}
 \mathrm{E}^{f}_{0}[\vec{\mathrm{R}}_{0}](\epsilon_{F}) = 
 \mathrm{E}^{\mathrm{tot}}_{0}[\vec{\mathrm{R}}_0] 
  - \mathrm{E}_{\mathrm{pristine}} - n_x \mu_x,
\end{equation}
where $\epsilon_{F}$ is the Fermi level, 
$\mathrm{E}^{\mathrm{tot}}_{0}[\vec{\mathrm{R}}_0]$ is the 
total energy of a system containing a neutral defect with atom positions at
$\vec{\mathrm{R}}_0$. $\mathrm{E}_{\mathrm{pristine}}$ represents the total energy of a pristine 
super cell of the same size. $n_x$ refers to the number of atoms of type $x$ 
added (positive) or removed (negative) from the pristine system, and $\mu_x$ 
is the atom's chemical potential \cite{PRB_KRP}. 
The neutral defect formation energy needs no electrostatic
correction term due to the absence of any long range electrostatic 
defect-defect interactions. 

The formation energy of a defect in charge state $q$ is given by \cite{PRB_KRP}:
\begin{equation}
 \label{eqn2}
  \mathrm{E}_{q}^{f}[\vec{\mathrm{R}}_{q}](\epsilon_{F}) = 
  \{\mathrm{E}^{\mathrm{tot}}_{q}[\vec{\mathrm{R}}_{q}] + \mathrm{E}^{\mathrm{corr}}_q \} 
  - \mathrm{E}_{\mathrm{pristine}}  + q \{ \epsilon_{\mathrm{vbm}}^{\mathrm{pristine}}+
  \epsilon_{F} - \Delta V_{0/p} \} - n_x \mu_x
\end{equation}

The first term on the right hand side is the total energy of a system containing
a defect in charge state q with the requisite finite-size electrostatic correction 
$\mathrm{E}^{\mathrm{corr}}_q$ (described below). The formation energy is now a function of the 
Fermi level in the system, $\epsilon_{F}$, with respect to the 
pristine valence band maximum (VBM), $\epsilon_{\mathrm{vbm}}^{\mathrm{pristine}}$. 
\begin{equation}
 \label{eqn_0p}
 \Delta V_{0/p} = V_0 |_{\mathrm{far}} - V_{p} 
\end{equation}
is a potential 
alignment term found by comparing the electrostatic potentials from a 
pristine calculation and far from the defect in a neutral defect calculation. 

\subsection{FNV correction scheme}
The electrostatic correction term, $\mathrm{E}^{\mathrm{corr}}_q$, is 
incorporated to correct the spurious interaction between the defect charge and 
its periodic images. In the FNV scheme, this term is given by \cite{PRB_KRP,PRL_FNV}:
\begin{equation}
 \label{eqn3}
  \mathrm{E}^{\mathrm{corr}}_q = \mathrm{E}^{\mathrm{lat}}_q - q\Delta V_{q-0/m}
\end{equation}
where $\mathrm{E}^{\mathrm{lat}}_q = \mathrm{E}^{\mathrm{iso,m}}_q - 
\mathrm{E}^{\mathrm{per,m}}_q$ is obtained from a model calculation. It 
involves solving the Poisson equation using a model charge distribution, 
$\rho^{m}(\mathbf{r})$, and model dielectric profile under periodic 
boundary conditions to obtain the potential, $V^{\mathrm{per,m}}_q(\mathbf{r})$ 
\cite{PRB_KRP,PRL_FNV}. 
$\mathrm{E}^{\mathrm{per,m}}_q$ is then given by:
\begin{equation}
 \label{eqn4}
  \mathrm{E}^{\mathrm{per,m}}_q = \frac{1}{2} \int_{\Omega} 
    \rho^{m}(\mathbf{r})V^{\mathrm{per,m}}_q(\mathbf{r})d\mathbf{r}
\end{equation}
where the integral is over the super cell volume, $\Omega$. $\mathrm{E}^{\mathrm{per,m}}_q$
is evaluated for larger super cells and extrapolated to obtain 
$\mathrm{E}^{\mathrm{iso,m}}_q$. Performing larger super cell calculations within 
this model is computationally inexpensive compared to a DFT calculation on such 
systems. $\mathrm{E}^{\mathrm{lat}}_q$ accounts for the long range interactions. 
The dependence of $\mathrm{E}^{\mathrm{corr}}_q$ on the model charge 
distribution is eliminated by the second term in \ref{eqn3}. $\Delta V_{q-0/m}$ is a 
potential alignment term found by comparing the model potential to the DFT difference 
potential: 
\begin{equation}
 \label{eqn5}
  \Delta V_{q-0/m} = (V_q^{\mathrm{DFT}} - V_0^{\mathrm{DFT}})|_{\mathrm{far}} - 
    V^{\mathrm{per,m}}_q |_{\mathrm{far}}
\end{equation}

\subsection{Model calculation}
The model super cell calculation for $\mathrm{E}^{\mathrm{per,m}}_q$ involves solving 
the Poisson equation for the periodic model potential, $V^{\mathrm{per,m}}_q(\mathbf{r})$
\cite{PRB_Noh,PRL_KP,PRB_Park}:
\begin{equation}
 \label{eqn6}
  \nabla.[\mathbf{\varepsilon}(\mathbf{r}) \nabla V^{\mathrm{per,m}}_q(\mathbf{r})]
  = - 4\pi \rho^m(\mathbf{r})
\end{equation}
where $\mathbf{\varepsilon}(\mathbf{r})$ is the dielectric tensor profile of the material,
$\rho^m(\mathbf{r})$ is the model charge distribution. The dielectric tensor can 
be obtained using Density Functional Perturbation Theory (DFPT) \cite{PRB_Gonze,RMP_Baroni}.
Eqn \ref{eqn6} can be effectively 
solved in the reciprocal space \cite{PRB_Noh,PRL_KP,PRB_Park}:

\begin{equation}
 \label{eqn7}
  \sum_{\mathbf{G'}} \sum_{i=1}^3 G_i G_i'\varepsilon_{ii}(\mathbf{G}-\mathbf{G'})
  V^{\mathrm{per,m}}_q(\mathbf{G'}) = 4\pi \rho^m(\mathbf{G})
\end{equation}
where $\varepsilon_{ii}$ are the diagonal terms of the dielectric tensor. 
$V^{\mathrm{per,m}}_q(\mathbf{G'=0})$ is set to zero, which is 
equivalent to introducing a uniform, neutralizing background charge. 
The number of \textbf{G} vectors used in the calculation is determined 
by an energy cut off. 

As one is interested in the long range corrections, $\varepsilon$ in the material
can be assumed to have no spatial profile; however, in general, it is a 3$\times$3 tensor. 
In a 3D bulk system with an isotropic dielectric, $V^{\mathrm{per,m}}_q$ can be obtained 
from Eqn \ref{eqn7} as:
\begin{equation}
 \label{eqn8}
  V^{\mathrm{per,m}}_q(\mathbf{G}) = \frac{4\pi \rho^m(\mathbf{G})}
    {\varepsilon|\mathbf{G}|^2}
\end{equation}

For slab or two-dimensional systems, on the other hand, the dielectric tensor, while 
not having any profile inside the material, would still have
a spatial profile in the aperiodic direction, say $z$ \cite{PRB_Noh,PRL_KP}. 
\begin{center}
  $\begin{bmatrix}
    \varepsilon^{1}_{\parallel}(z) & 0.0 & 0.0 \\
    0.0 & \varepsilon^{2}_{\parallel}(z) & 0.0 \\
    0.0 & 0.0 &  \varepsilon_{\perp}(z)\\
  \end{bmatrix}$
\end{center}
where $\varepsilon^{1}_{\parallel}(z)$ and $\varepsilon^{2}_{\parallel}(z)$ 
is the dielectric profile for the in-plane dielectric constants $\vec{a}_1$ and $\vec{a}_2$
and $\varepsilon_{\perp}(z)$ is the dielectric profile for the out-of-plane dielectric
constant of the material.

\begin{figure}
 \centering
 \includegraphics[scale=0.3]{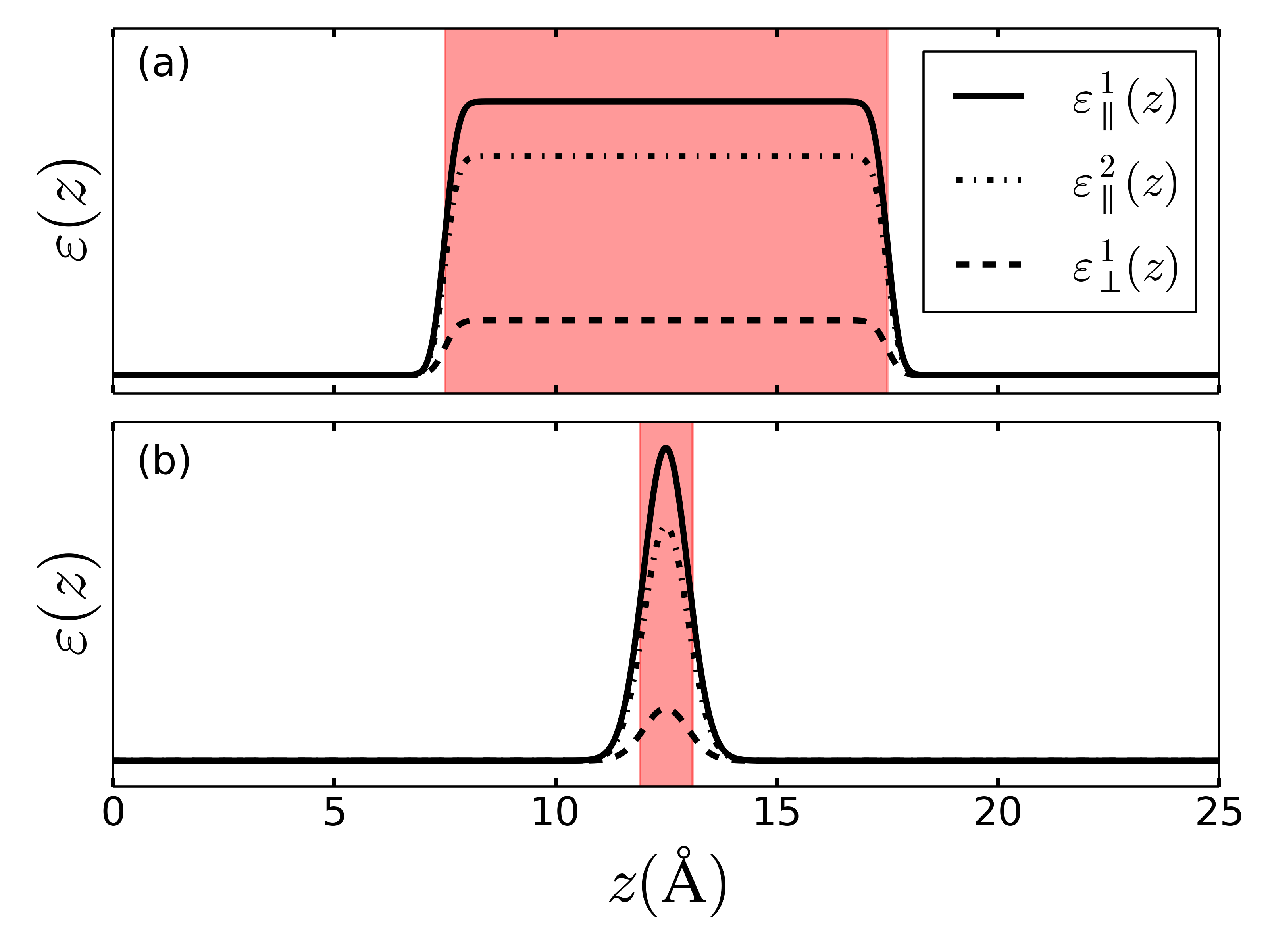}
 \caption {\label{fig1} (Color online) (a) Sample profile for slab systems or
 2D materials with more than one atom thickness. (b) Sample profile for single 
 atom thickness 2D materials. }
\end{figure}

For slab systems or a 2D material with more than one atom thickness, like 
transition metal dichalcogenides, phosphorene, etc., the dielectric profile used is
of the form, \cite{PRB_Noh,PRL_KP}:
\begin{equation}
 \label{eqn9}
  \varepsilon(z) = \frac{1}{2}(\varepsilon^v-\varepsilon^m)
    \mathrm{erf}(\frac{z - t_1}{s}) - \frac{1}{2}(\varepsilon^v-\varepsilon^m)
    \mathrm{erf}(\frac{z - t_2}{s}) + \varepsilon^v
\end{equation}
where $t_1$ and $t_2$ are the edges of the slab in the $z$-direction in the 
simulation cell. A parameter, $s$, is used with the error 
function (erf) to smoothen the profile at the slab edges. $\varepsilon^m$ 
is the dielectric constant of the material, found using DFPT \cite{PRB_Noh}.
$\varepsilon^v$ is the dielectric constant of the space surrounding the material, set to 
1 for vacuum. A sample slab profile is shown in Fig \ref{fig1} (a).
For a 2D material with one atom thickness, like Boron Nitride (BN), 
Silicon Carbide (SiC), graphene, etc., the 
dielectric profile used is a gaussian to mimic the planar-averaged charge 
density of the material in the $z$-direction. A sample profile is shown in 
Fig \ref{fig1} (b). 

For slab or 2D systems, Eqn \ref{eqn7} can then be written as:
\begin{multline}
 \label{eqn10}
  \sum\limits_{G_z^\prime}[\varepsilon_\perp (G_z - G_z^\prime)] G_z^\prime G_z V^{\mathrm{per,m}}_q(G_x,G_y,G_z^\prime)   \\
  + \sum\limits_{G_z^\prime}[\varepsilon_\parallel (G_z - G_z^\prime)] 
  (G_x^2 + G_y^2) V^{\mathrm{per,m}}_q(G_x, G_y, G_z^\prime) = 4\pi\rho^m(G_x, G_y, G_z)
\end{multline}
The average potential, 
$V^{\mathrm{per,m}}_q(\mathbf{G}=0)$, is set to zero, to introduce a neutralizing 
bakground charge.

\begin{figure}
 \centering
 \includegraphics[scale=0.3]{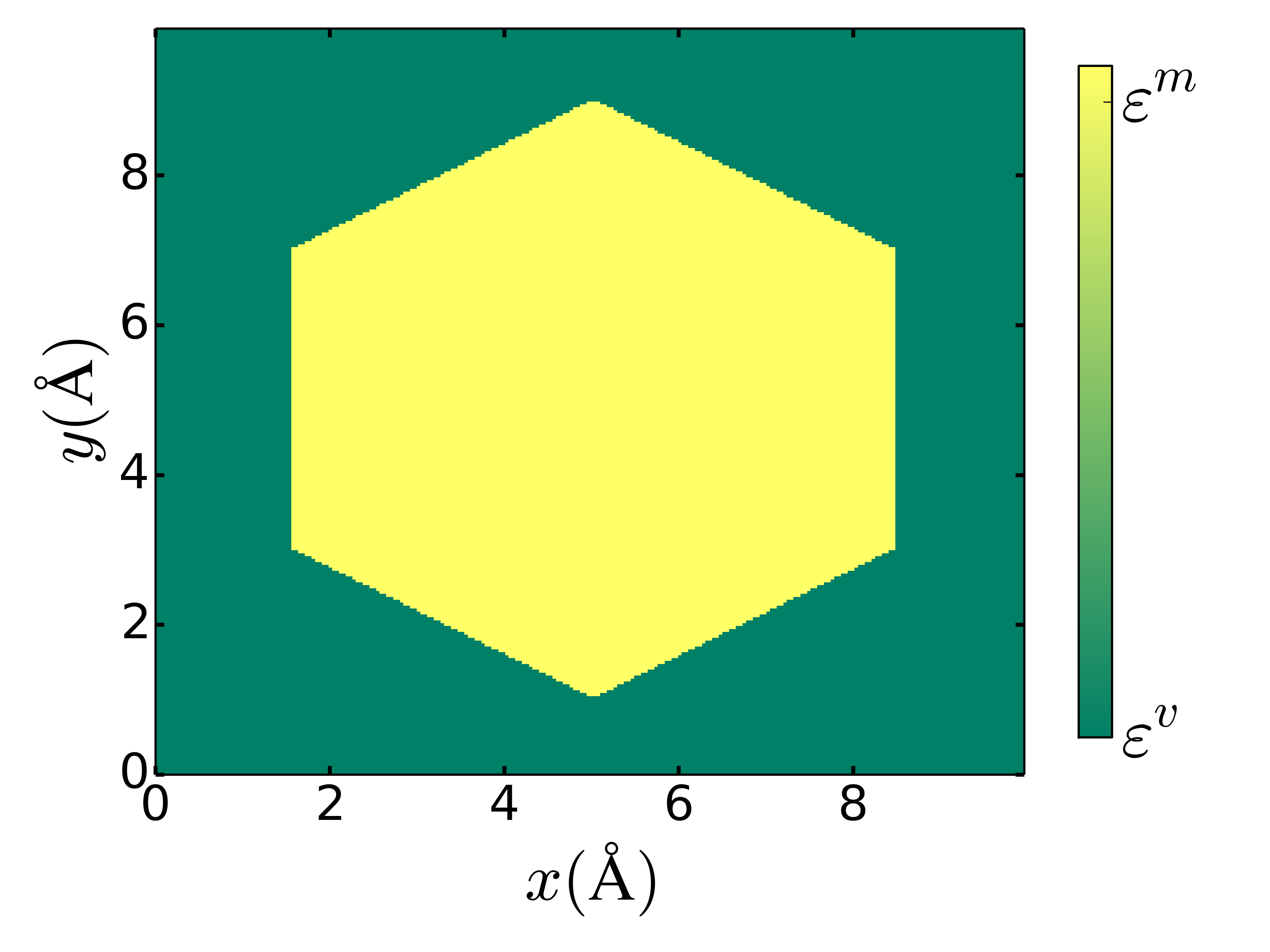}
 \caption {\label{fig2} (Color online) Sample profile for a nanowire 
 along $z$ direction  with a hexagonal cross-section. $\varepsilon^m$ is 
 the dielectric constant of the material and $\varepsilon^v$ that of 
 the environment ($\varepsilon^v = 1$ for vacuum).}
\end{figure}
\begin{figure}
 \centering
 \includegraphics[scale=0.3]{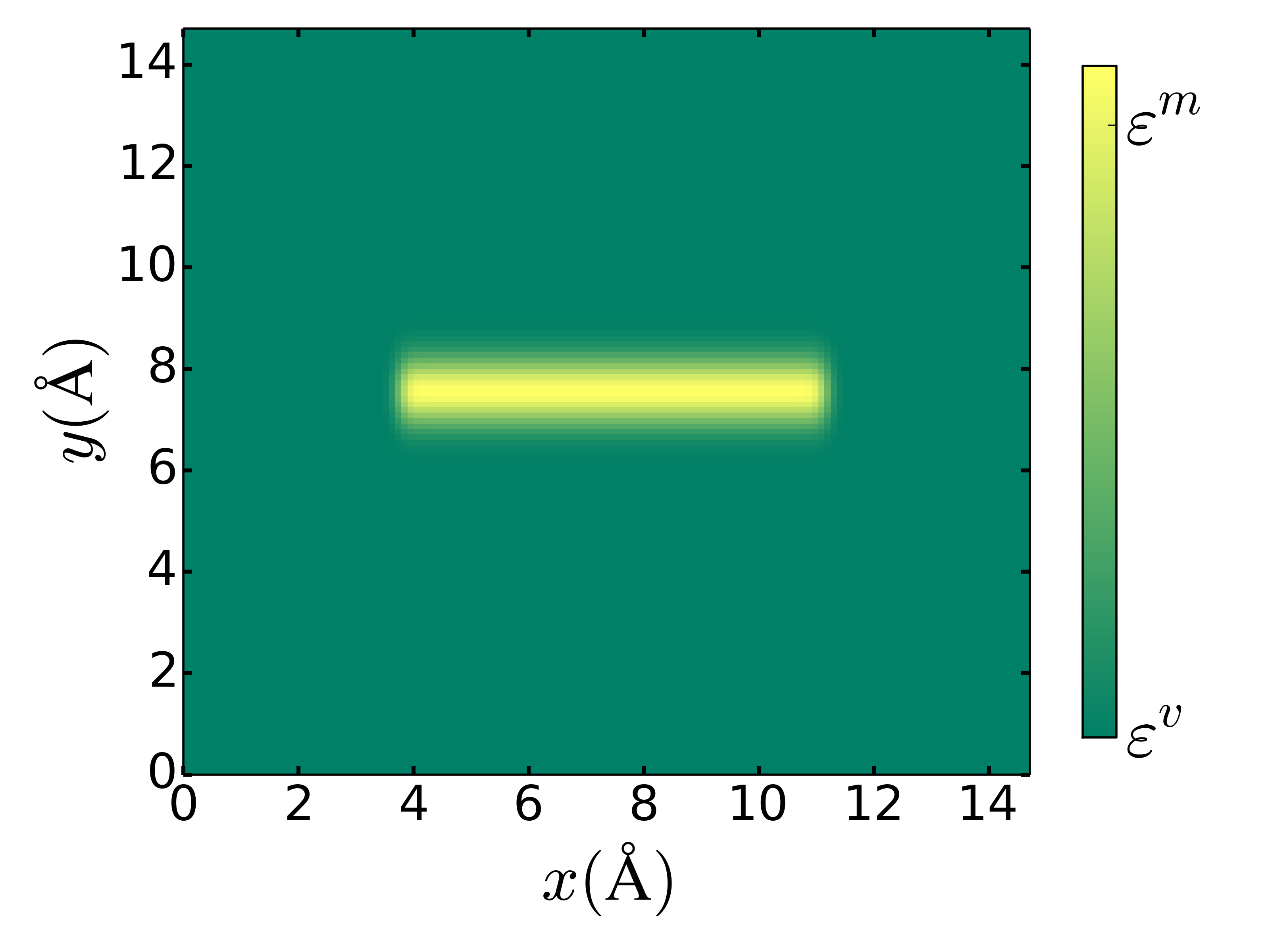}
 \caption {\label{fig3} (Color online) Sample profile for a nanoribbon
 with periodicity along the $z$ direction. $\varepsilon^m$ is
 the dielectric constant of the material and $\varepsilon^v$ that of
 the environment ($\varepsilon^v = 1$ for vacuum).}
\end{figure}

For a quasi-1D system like nanowires or nanoribbons, with perodicity 
along the $z$ direction, the dielectric tensor profile is of the form
\cite{PRB_Park} :
\begin{center}
  $\begin{bmatrix}
    \varepsilon_{\perp}(x,y) & 0.0 & 0.0 \\
    0.0 & \varepsilon_{\perp}(x,y) & 0.0 \\
    0.0 & 0.0 &  \varepsilon_{\parallel}(x,y)\\
  \end{bmatrix}$
\end{center}
where $\varepsilon_{\perp}(x,y)$ is the profile for the dielectric constant 
perpendicular to the wire and $\varepsilon_{\parallel}(x,y)$ is the profile for 
the dielectric constant along the wire. A sample profile for a silicon nanowire 
along the [111] direction is shown in Fig. \ref{fig2}. The wire, oriented 
along the $z$ direction, has a hexagonal cross-section \cite{PRB_Park}. 
The profile is constructed such that the points inside the wire 
have the dielectric constant of the material, $\varepsilon^m$, and 
the points outside have $\varepsilon^v$. A sample profile for a BN nanoribbon 
is shown in Fig. \ref{fig3}. The profile is constructed to mimic the electron 
charge density of the material by combining a slab like profile (Eqn \ref{eqn9})
along the width of the nanoribbon and a gaussian along the out-of-plane direction. 
Eqn \ref{eqn7}, for 1D systems, can then be written as \cite{PRB_Park}:

\begin{multline}
  \label{eqn_n1}
  \sum\limits_{G_x^\prime,G_y^\prime}[\epsilon_\parallel(G_x - 
  G_x^\prime,G_y - G_y^\prime) G_z^2  \\  + \epsilon_\perp (G_x - 
  G_x^\prime,G_y - G_y^\prime) (G_x G_x^\prime + G_y G_y^\prime) 
  ]V(G_x^\prime, G_y^\prime, G_z) \\= 4\pi\rho(G_x, G_y, G_z) 
\end{multline}
We set $V(\mathbf{G} = 0) = 0$ here, to introduce a neutralizing 
background charge.

\subsection{Defect eigenvalue correction}
The defect eigenvalues, like the formation energy, show slow convergence with 
the super cell size \cite{JPCM_Chen,PRB_CP}. The correction to the eigenvalue is 
given by \cite{PRB_CP}:
\begin{equation}
  \label{eqn_n2}
  \epsilon_{q}^{\mathrm{corr}} = \frac{-2}{q} \mathrm{E}_q^{\mathrm{corr}}
\end{equation}
where  $\mathrm{E}_q^{\mathrm{corr}}$ is the FNV correction term, Eqn \ref{eqn3}. The 
correction is negative for positively charged defects and positive for 
negatively charged defects. Defect level corrections are crucial in studying 
defects using the DFT+GW formalism \cite{JPCM_Chen,PRB_CP,PRB_Andrei,PRL_Jain, RMP_Walle} and 
also in interpreting absorption experiments.

\section{Test Systems}
We show the performance of our code in correcting the formation energies and 
defect levels in the following systems: 
\begin{enumerate}
  \item Bulk: Carbon vacancy in diamond in charge state -2, $V_{C}^{-2}$.
  \item 2D: Sulfur vacancy in monolayer MoS$_2$ in charge state -1, $V_{S}^{-1}$.
  \item 1D: Boron vacancy in BN nanoribbon in charge state -1, $V_{B}^{-1}$.
\end{enumerate}

All the DFT calculations are performed using the plane-wave, pseudopotential 
package, Quantum Espresso \cite{QE.Giannozi}. We 
perform simulations on a range of super cell sizes and compare the extrapolated 
formation energy and defect level (to the isolated limit) with the corrected 
formation energy and defect level.

\subsection{Vacancy in diamond, $V_{C}^{-2}$}
We perform DFT calculations on different cubic super cell sizes with number of C atoms 
ranging from 64 to 1024. We use norm-conserving pseudopotentials and the LDA 
exchange correlation functional \cite{PRB.Perdew}. The wavefunctions are expanded in plane-waves upto 
an energy cut off of 60 Ry. For the 64 atom, 2$\times$2$\times$2 super cell, 
a k-point sampling of 5$\times$5$\times$5 was used. An equivalent sampling is used 
for the other super cell sizes. We did not include any atomic relaxations. 

\begin{figure}
 \centering
 \includegraphics[scale=0.22]{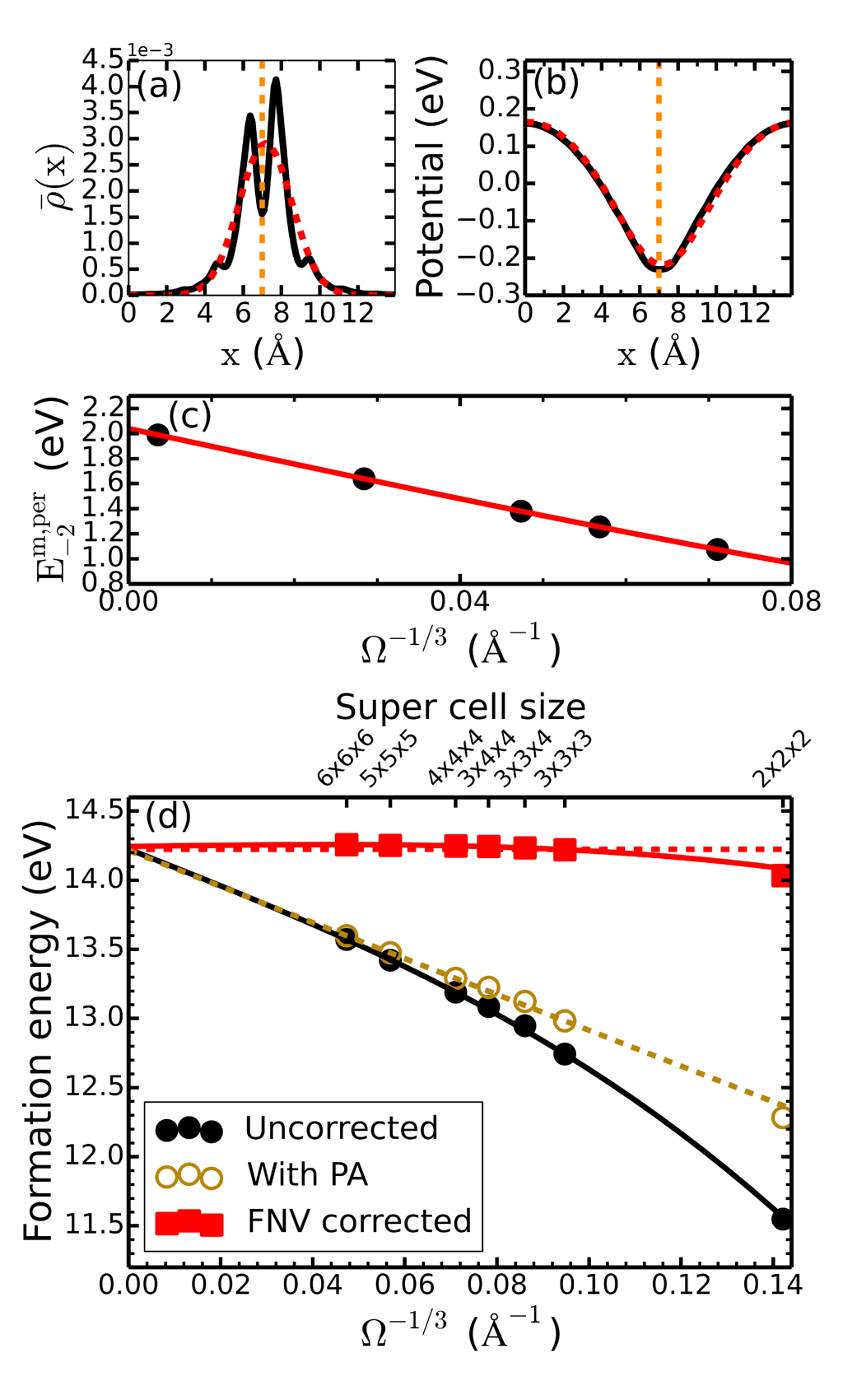}
 \caption {\label{fig4} (Color online) 
 (a) Black solid line refers to the planar averaged defect wavefunction 
     charge density. 
     Red dashed line refers to the planar averaged model charge 
     density.
 (b) Black solid line refers to the planar averaged DFT difference 
     potential, $V_{q}^{\mathrm{DFT}} - V_{0}^{\mathrm{DFT}}$, in \ref{eqn5} 
     from a 3$\times$3$\times$3 super cell.
     Red dashed line refers to the planar averaged model potential for 
     the same size super cell.
 (c) Scaling of the total energy from the model calculation with super cell 
     volume, $\Omega$.
 (d) Black solid line shows the fit to the scaling of the uncorrected 
     formation energy (filled black dots) with super cell size. The unfilled 
     golden circles show the formation energy with only potential alignment 
     corrections. The red squares show the FNV corrected formation energy for 
     various super cell sizes. The red solid line is a fit to the FNV corrected 
     formation energies and the horizontal red dashed line marks the extrapolated
     value of the uncorrected formation energies. 
          }
\end{figure}

\begin{figure}
 \centering
 \includegraphics[scale=0.3]{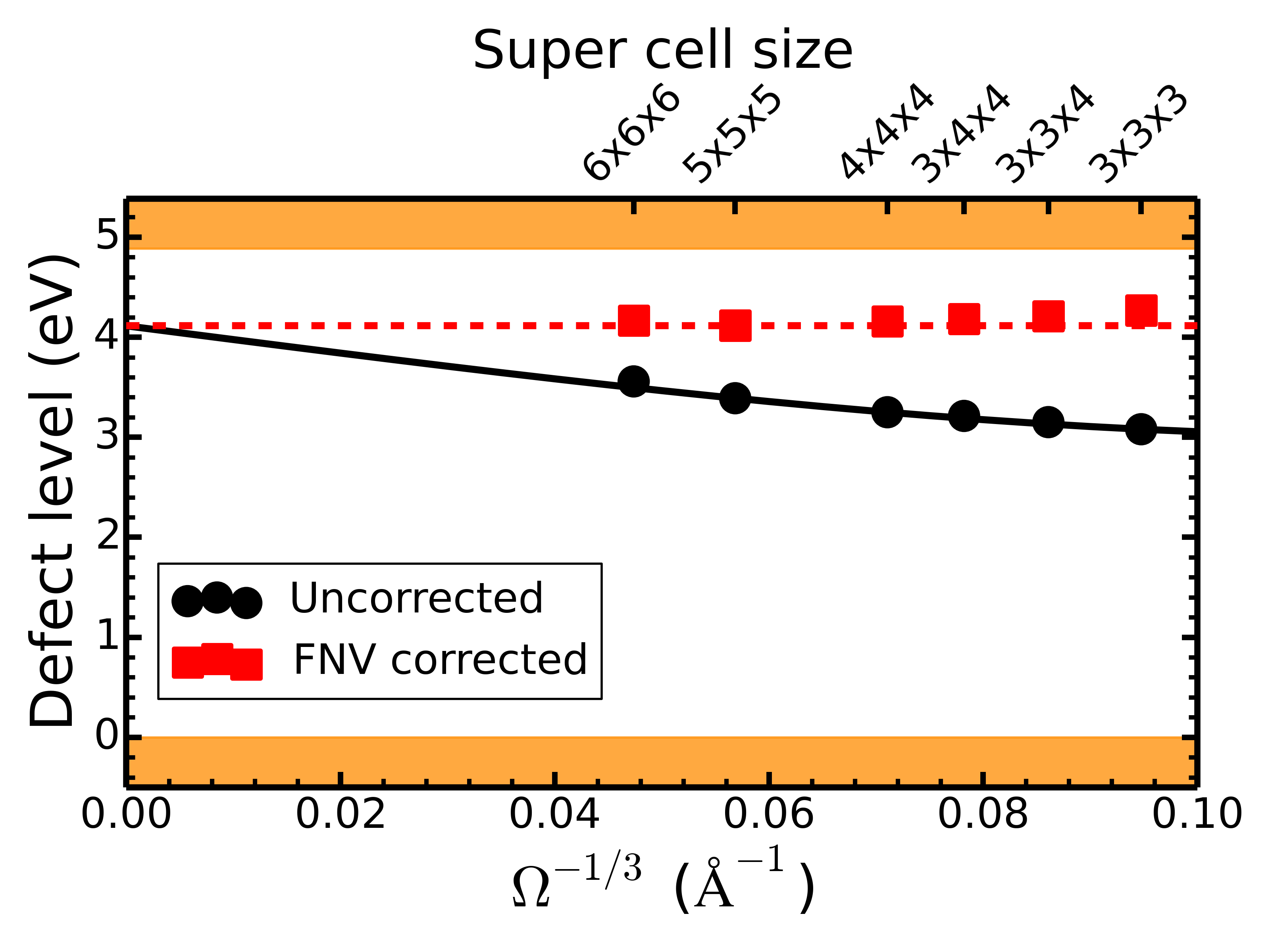}
 \caption {\label{fig5} (Color online) 
  Black solid line shows the fit to the scaling of the uncorrected
  defect eigenvalue (filled black dots) with super cell size. The horizontal 
  red dashed line marks the extrapolated value of the uncorrected 
  defect level. The red squares show the FNV corrected defect eigenvalue. 
          }
\end{figure}

The model calculation is performed using the CoFFEE code, solving Eqn \ref{eqn8}.
The defect is modelled with a gaussian of width 2.6 bohr, a dielectric constant of 5.76
and a plane wave energy cut off of 16 Ry. Fig. \ref{fig4} (a) 
shows the defect wavefunction charge density, $|\psi_{d}|^2$, and 
the model charge density. For bulk systems, it is not necessary
that the width of the Gaussian model charge match the defect 
wavefunction charge density. It suffices if the width is appropriately
small to keep the model charge inside the cell. Note also that the plane wave 
energy cut off required to converge the model calculation is inversely
proportional 
to the width of the Gaussian. 
Fig. \ref{fig4} (c) shows the extrapolation of the model energy,
$\mathrm{E}^{\mathrm{per,m}}_{-2}$ (Eqn \ref{eqn4}), with 
super cell size, $\Omega$.
The energy is fit with a polynomial of the form: 
\begin{equation}
\label{eqn11}
p(\Omega) = f_1 + f_2/\Omega^{1/3} + f_3/\Omega 
\end{equation}
$f_1$ here corresponds to the isolated model energy, $\mathrm{E}^{\mathrm{iso,m}}_{-2}$. 
The fitting parameters here are: $f_1 = 2.04$, $f_2 = -14.19 $ and $f_3 = 120.85$. 
The lattice correction is then given by: 
$\mathrm{E}^{\mathrm{iso,m}}_{-2} - \mathrm{E}^{\mathrm{per,m}}_{-2}$ (Eqn \ref{eqn3}). 
Fig. \ref{fig4} (b) compares the planar averaged model potential and the DFT difference 
potential which contribute to the potential alignment term, 
$\Delta V_{q-0/m}$ (Eqn \ref{eqn5}) for a 3$\times$3$\times$3 super cell. 
The defect in this calculation is at the center of the cell. The model 
potential matches well with the DFT difference potential far from the 
defect, leading to a very small correction. On the other hand, the
potential alignment term, $\Delta V_{0/p}$ in Eqn \ref{eqn2}, which compares the potential 
far from the neutral defect to the pristine, has a substantial contribution.

The uncorrected formation energy of $V_C^{-2}$ for various supercell sizes is 
fit using Eqn \ref{eqn11} to extrapolate to the isolated limit as shown 
in Fig. \ref{fig4} (d). The chemical potential of carbon is taken from bulk diamond.
The FNV corrected formation energy 
shows good agreement with the extrapolated 
value for super cell sizes larger than 2$\times$2$\times$2. The correction scheme 
performs well when the defect wavefunction is localized well within the super cell; 
this is not the case with the 2$\times$2$\times$2 super cell. 
The defect level accomodating the -2 charge is shown in Fig. \ref{fig5} with respect to the 
pristine VBM eigenvalue in the charged super cell. It shows a similar scaling with 
super cell size and is corrected using Eqn \ref{eqn_n2}.

\subsection{Vacancy in monolayer MoS$_2$, $V_{S}^{-1}$}
We study sulfur vacancies in monolayer MoS$_2$ to demonstrate the electrostatic 
corrections in 2D or slab systems. The DFT calculations are performed using 
PAW pseudoptentials \cite{PRB.PAW}, PBE scheme for the exchange correlation functional \cite{JCP.PBE}. 
A wavefunction cut off of 50 Ry, and charge density cut off of 500 Ry is used. 
We perform calculations on four super cell sizes: 
$\alpha \times \alpha \times \alpha$, for $\alpha$ = 4, 5, 6 and 8. 
The  $\alpha$ refers to the scaling of the cell dimension.
It has been reported that scaling just the in-plane 
supercell, keeping the amount of vacuum fixed, leads to a divergence in the 
model energy \cite{PRL_KP}. 
We hence uniformly scale the vacuum with 
the in-plane super cell. 
For $\alpha = 6$, the vacuum thickness is $\sim 16 \mathrm{\AA}$ and
thickness of MoS$_2$ is $3.2 \mathrm{\AA}$.
The k-point sampling for these is chosen to effectively sample the unit 
cell Brillouin zone with a grid better than 12$\times$12$\times$1. 
The atoms are relaxed to their equilibrium positions such that the 
force on each atom is  $\leq 10^{-2}$ eV/\AA.

\begin{figure}
 \centering
 \includegraphics[scale=0.3]{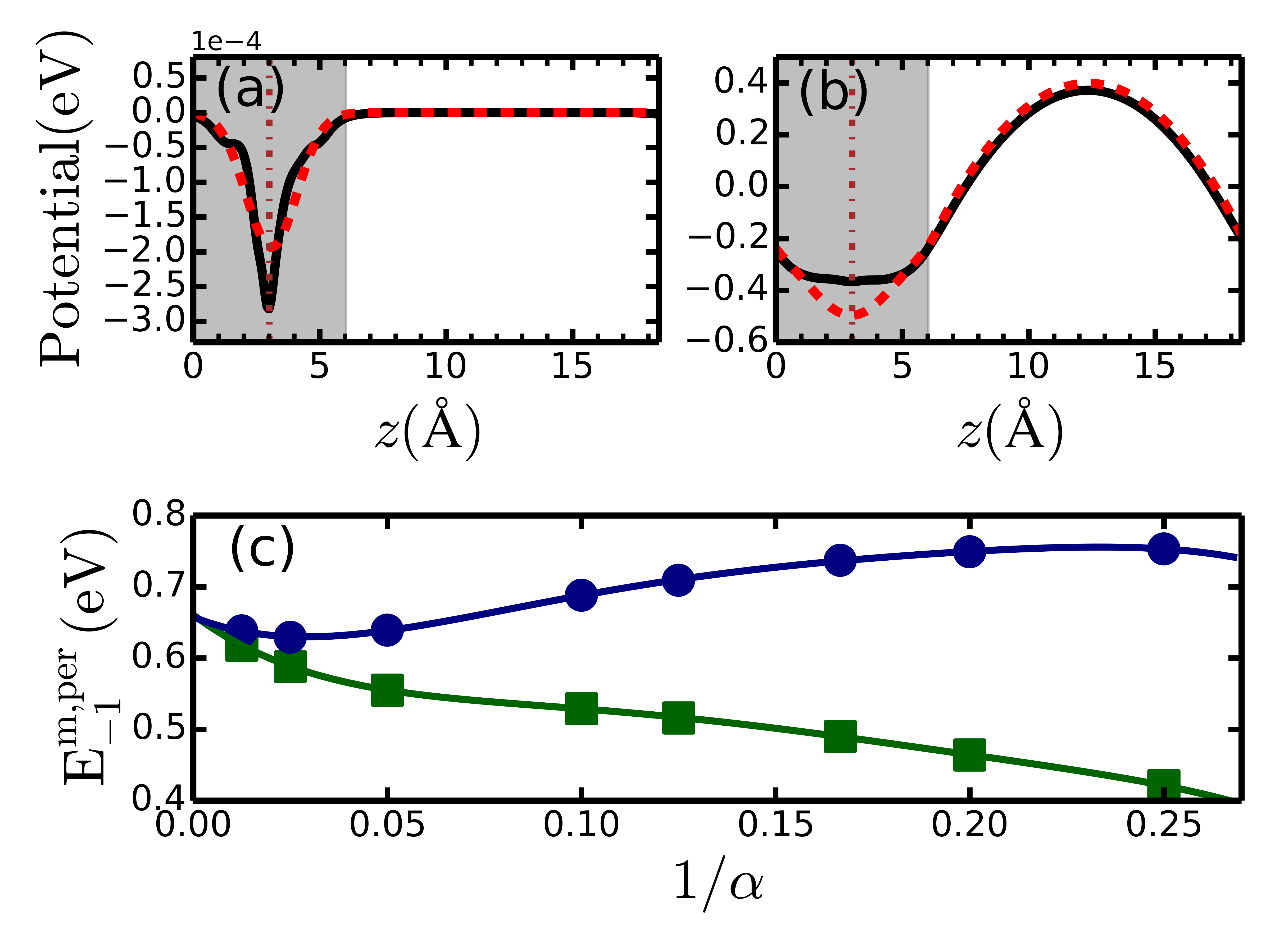}
 \caption {\label{MoS2-fig1} (Color online)
  (a) Black solid line shows the planar averaged defect wavefunction charge 
      density, along the out-of-plane direction, for the $\alpha = 6$ 
      super cell of MoS$_2$. The 
      red dashed line shows the planar averaged Gaussian model charge. 
      The shaded gray region marks the location and width of the dielectric 
      profile (Eqn. \ref{eqn9}). 
  (b) Black solid line shows the planar averaged DFT difference potential: 
      $V_{-1}^{\mathrm{DFT}} - V_{0}^{\mathrm{DFT}}$, along the 
      out-of-plane direction for the same cell.
      The red dashed line shows the planar averaged model potential. 
      The vertical brown dot-dashed line marks the center of the Gaussian model 
      charge.
  (c) Total energy from the model calculation with two super cell size scalings: 
      $\alpha \times \alpha \times \alpha$ (green squares) and 
      $\alpha \times \alpha \times1.6 \alpha$ (blue circles).
  }
\end{figure}

The model calculation is performed solving Eqn \ref{eqn10}. The dielectric profile is 
chosen as in Eqn \ref{eqn9}, with $\varepsilon_m = 15$ in the direction parallel to 
the material and $\varepsilon_m = 2$ in the perpendicular direction. 
The smoothening parameter, s, is chosen to be 0.38 bohr. 
The dielectric constants are determined using DFPT as 
prescribed in reference \cite{PRB_Noh}. 
The plane wave energy cut off 
is set to 30.0 Ryd. 
The defect is modelled with a Gaussian of
width 1.9 bohr as shown in Fig. \ref{MoS2-fig1} (a). 
It is important here to choose a Gaussian width that mimics 
the defect charge density. Choosing a width too large that spills 
beyond the gray shaded region in Fig. \ref{MoS2-fig1} (a) would affect the 
results. The defect wavefunction charge density can be fit to a
Gaussian using the script \emph{g\_fit.py} supplied with the CoFFEE code. 

Fig \ref{MoS2-fig1} (b) compares the planar averaged  model potential with the 
planar averaged DFT difference potential, $V_{-1}^{\mathrm{DFT}} - V_{0}^{\mathrm{DFT}}$,
for the $\alpha = 6$ supercell. The DFT difference potential used here is from 
unrelaxed calculations and shows good agreement with the model potential 
far from the defect. 
Fig \ref{MoS2-fig1} (c) shows the variation of the model electrostatic 
energy with two different scalings of the super cell size: $\alpha \times \alpha \times \alpha$ 
and $\alpha \times \alpha \times 1.6 \alpha$. 
For $\alpha \times \alpha \times 1.6 \alpha$ super cell scaling, 
the vacuum thickness is 1.6 times larger than that 
of $\alpha \times \alpha \times \alpha$ scaling.
The interaction between the periodic charges is unscreened in the out-of-plane
direction, and screened by the material in the in-plane direction. 
The model energy is thus strongly dependent on the vacuum thickness, 
for small super cell sizes. 
As expected, at the infinite limit ($\alpha = \infty$), 
the two scalings extrapolate to the same value. 
The model scaling plots clearly do not follow a simple fit as 
in the case of bulk solids due to the spatially varying
dielectric profile.
The energy is 
fit with a polynomial of degree 
five in $\alpha^{-1}$: $\mathrm{E}^{\mathrm{per,m}}_{-1}(\alpha) = 
f_0 + f_1 \alpha^{-1} + f_2 \alpha^{-2} + f_3 \alpha^{-3} 
+ f_4 \alpha^{-4} + f_5 \alpha^{-5}$ \cite{PRB_Noh}. 
The isolated model energy is given by $f_0 = \mathrm{E}^{\mathrm{iso,m}}_{-1}$,
 and the lattice correction is given by:
$\mathrm{E}^{\mathrm{iso,m}}_{-1} - \mathrm{E}^{\mathrm{per,m}}_{-1}$. The fitting
parameters, for $\alpha \times \alpha \times \alpha$ scaling,
are found to be: $f_0 = 0.66$, $f_1 = -4.04$, $f_2 = 54.95$, 
$f_3 = -388.69$, $f_4 = 1267.38$ and $f_5 = -1579.27$. 

\begin{figure}
 \centering
 \includegraphics[scale=0.3]{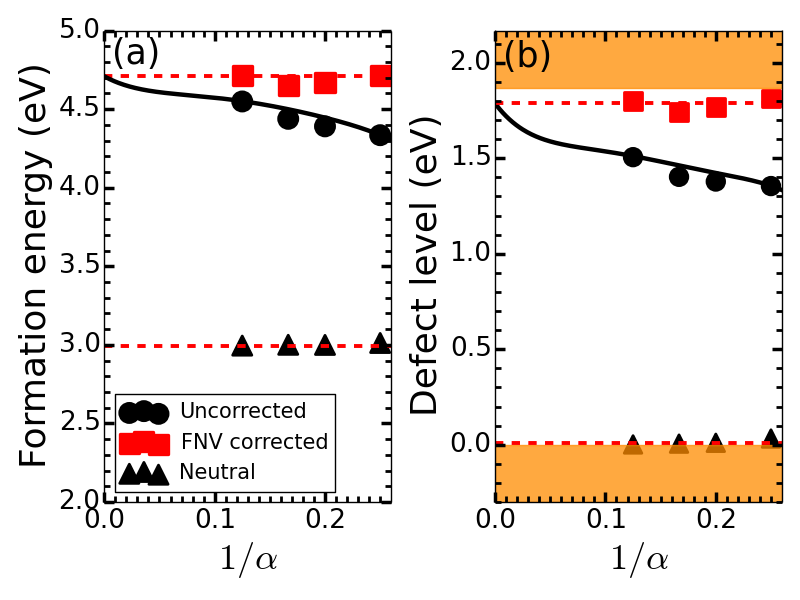}
 \caption {\label{MoS2-fig2} (Color online)
  (a) Black solid line shows the fit to the scaling of the uncorrected 
      formation energy (filled black dots) of $V_{S}^{-1}$. The red squares show 
      the FNV corrected formation energy, the horizontal red dashed line shows 
      the exptrapolated value from the uncorrected formation energies. The traingles 
      show the formation energy of the neutral defect.
  (b) Black solid line shows the fit to the uncorrected defect level (filled 
      black dots) holding the added electron in $V_{S}^{-1}$. The red squares 
      show the FNV corrected defect level. The horizontal red dashed line shows
      the exptrapolated value from the uncorrected levels. The triangles show 
      the position of the filled defect level in $V_{S}^{0}$. The orange bands 
      mark the valence and conduction band edges, the VBM is set to zero.
  }
\end{figure}

Fig \ref{MoS2-fig2} (a) shows the scaling of the uncorrected formation energy 
of neutral S vacancy, $V_S^{0}$, and -1 charged S vacancy, $V_S^{-1}$, as a 
function of super cell size. The chemical potential of sulfur is taken from
the cyclo-S$_8$ allotrope of sulfur. 
The neutral vacancy formation energy is 
well converged and shows no scaling with system size. The scaling of the charged 
vacancy formation energy is fit with a polynomial of the form \cite{PRB_Noh}: 
$\mathrm{E}_{-1}^{f} (\alpha) = 
t_0 + (f_1 \alpha^{-1} + f_2 \alpha^{-2} + f_3 \alpha^{-3} + f_4 \alpha^{-4}
+ f_5 \alpha^{-5})  + t_3 \alpha^{-3}$. The additional terms are necessary to capture the 
scaling of the potential alignment term with system size.
$t_0$ and $t_3$ are found to be 4.71 and -8.69 respectively. The FNV corrected formation
energy shows excellent agreement with the extrapolated value for all the super cell
sizes considered. Fig \ref{MoS2-fig2} (b) shows the dependence of the defect level
accomodating the -1 charge with system size. The uncorrected values are fit with a
polynomial of the form:
$\epsilon_{-1}(\alpha) = t_0 + 2\times(f_1 \alpha^{-1} + f_2 \alpha^{-2} 
+ f_3 \alpha^{-3} + f_4 \alpha^{-4} + f_5 \alpha^{-5}) + t_3 \alpha^{-3}$.
The corrected eigenvalues are in good agreement with the extrapolated value.

\subsection{Vacancy in BN nanoribbon, $V_{B}^{-1}$}
\begin{figure}
 \centering
 \includegraphics[scale=0.3]{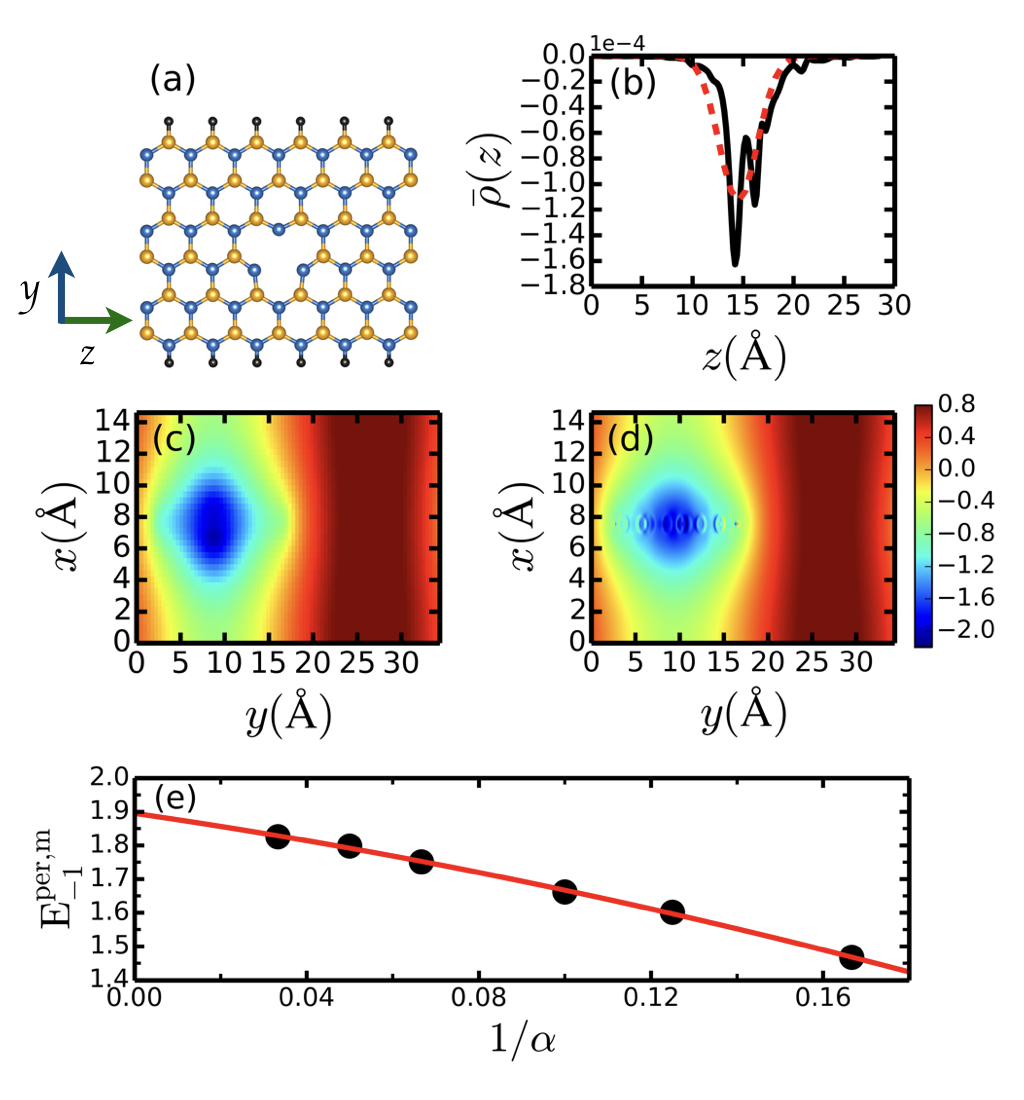}
 \caption {\label{NR-fig1} (Color online)
  (a) B vacancy in the 6$\times$19$\times$6 BN nanoribbon super cell. The yellow
      spheres denote the B atoms, blue spheres the N atoms and black spheres the
      H atoms.
  (b) Black solid line refers to the planar averaged defect wavefunction charge    
      density, plotted along the $z$ direction. The red dashed lines refers to the
      planar averaged model Gaussian charge density. 
  (b) scaling of the energy from the model calculation for $V_{B}^{-1}$
      with super cell size, $\alpha$ (1$\times$2.31$\times$1).
  (c) $z$-averaged model potential for the 6$\times$19$\times$6 cell.
  (d) $z$-averaged DFT difference potential: $V_{-1}^{\mathrm{DFT}} - V_0^{\mathrm{DFT}}$
      for the 6$\times$19$\times$6 cell.
  }
\end{figure}

To show the application of the code for 1D systems, we study B vacancy in a BN nanoribbon 
of thickness 13.39 \AA \cite{PE_Mota}. We perform calculations on three super cell sizes: 
$\alpha$ (1$\times$2.31$\times$1), for $\alpha = 6,8,10$. 
For $\alpha=6$, the simulation cell dimensions are 15$\mathrm{\AA}$, 
35$\mathrm{\AA}$ and 15$\mathrm{\AA}$, in the $x$, $y$ and $z$ directions 
repectively. The $z$ direction is the periodic direction. A vacuum of 
15$\mathrm{\AA}$ and 22$\mathrm{\AA}$ has been introduced 
in the out-of-plane $x$ direction and lateral $y$ direction. 
The thickness of the 
ribbon is fixed in these cell sizes, with the number of atoms in the periodic direction 
and the vacuum padding scaling with $\alpha$. The ribbon is passivated 
on either side with H atoms.
Ribbon in $\alpha = 6$ super cell with a B vacancy is shown in Fig \ref{NR-fig1} (a). 
The DFT calculations are performed using norm conserving pseudopotentials and
the PBE scheme is used for the exchange correlation functional \cite{JCP.PBE}. 
A plane wave energy cut off of 70 Ry is used for the wavefunctions. The Brillouin 
zone is sampled with a 1$\times$1$\times$2 grid for $\alpha = 6,8$ and with the 
$\Gamma$ point for $\alpha = 10$. We relaxed the atoms in the super cell containing the 
defect until the force on each atom is  $\leq 10^{-2}$ eV/\AA.

For the model calculation, we use a dielectric profile 
as shown in Fig \ref{fig3}. A slab like profile, similar to the one used for MoS$_2$, 
is used along the width of the nanoribbon and a gaussian profile in the out-of-plane, 
$x$ direction. The width of the slab is taken to be that of the width of the 
nanoribbon, with a smoothening parameter of 0.37 bohr. The gaussian width is taken to 
be 0.86 bohr, chosen to mimic the electron charge density in that direction. 
$\epsilon^m$ used is 2.9 along the $x$ direction and 12 along the other 
two directions. These are computed from DFPT calculations on a monolayer BN sheet. 
We use a gaussian model charge with width 1.89 bohr and integrated charge -1. 
The planar averaged model charge and the defect wavefunction charge density are 
shown in Fig. \ref{NR-fig1} (b). 
The calculation is performed with a plane wave energy cut off of 16.0 Ry. 
Fig \ref{NR-fig1} (e) shows the scaling of the computed 
$\mathrm{E}^{\mathrm{per,m}}_{-1}$ as a function of 1/$\alpha$. 
The points are fit with a third degree polynomial in $\alpha^{-1}$: 
$\mathrm{E}^{\mathrm{per,m}}_{-1}(\alpha) = 
f_0 + f_1 \alpha^{-1} + f_2 \alpha^{-2} + f_3 \alpha^{-3} $.
$f_0$ then corresponds to the isolated model energy, $\mathrm{E}^{\mathrm{iso,m}}_{-1}$
, and the lattice correction is given by: 
$\mathrm{E}^{\mathrm{iso,m}}_{-1} - \mathrm{E}^{\mathrm{per,m}}_{-1}$. The fitting 
parameters are found to be: $f_0 = 1.90$, $f_1 = -1.85$, $f_2 = -4.35$ and $f_3 = 0.58$. 
The potential alignment terms, $\Delta V_{q-0/m}$ and  $\Delta V_{0/p}$, are calculated 
by comparing the potentials in the out-of-plane, $x$ direction, far from the defect.
Both these terms are found to be small ($\textless$ 20 meV).
Fig \ref{NR-fig1} (c) and (d) show the DFT difference potential in Eqn \ref{eqn5} and 
the model potential, respectively, for $\alpha = 6$ super cell size. The defect is at 
(x, y) = (7.5, 9) in the figure. The model potential matches well with the DFT difference 
potential far from the defect.

\begin{figure}
 \centering
 \includegraphics[scale=0.3]{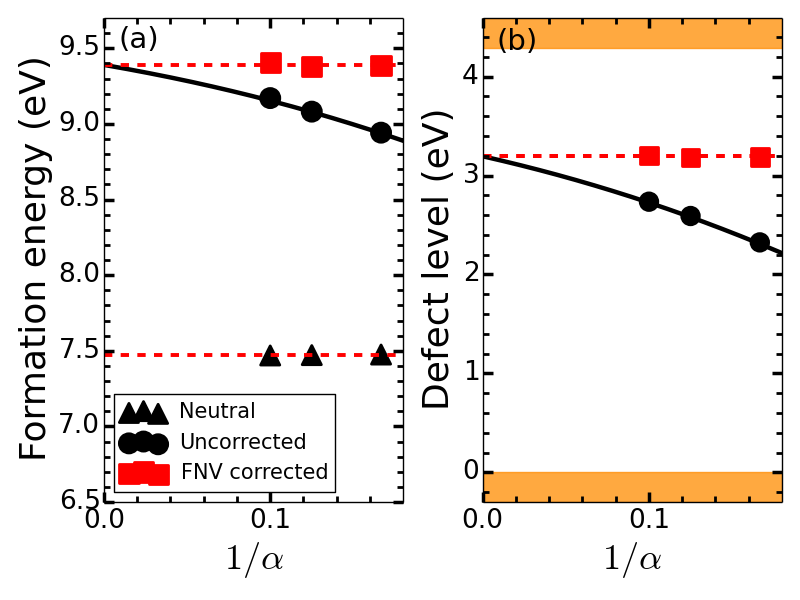}
 \caption {\label{NR-fig2} (Color online) 
 (a) Black solid line shows the fit to the scaling of the uncorrected formation 
     energy of -1 charged B vacancy in BN nanoribbon (filled black dots) with super 
     cell size. The red squares show the FNV corrected formation energy. The 
     horizontal red dashed line shows the extrapolated value from the uncorrected 
     formation energies. The triangles show the formation energy of the neutral defect. 
 (b) Filled black dots show the uncorrected defect eigenvalue in the gap as a function of 
     system size. Black solid line is the fit to these, and the isolated limit 
     extrapolation is marked with the red dashed line. The red squares mark the 
     FNV corrected eigenvalues. The orange bands
      mark the valence and conduction band edges, the VBM is set to zero.
  }
\end{figure}

Fig \ref{NR-fig2} (a) shows the scaling of the uncorrected formation energy of 
neutral vacancy, $V_B^{0}$, and -1 charged vacancy, $V_B^{-1}$, as 
a function of $1/\alpha$. The formation energy is computed 
for nitrogen-rich conditions. The chemical potential for N is 
taken from N$_2$ molecule.
Formation energy of the neutral vacancy is well 
converged and shows no scaling with the system size. The uncorrected formation 
energy of the charged vacancy is fit with a polynomial of the form: 
$\mathrm{E}_{-1}^{f} (\alpha) = t_0 + (f_1 \alpha^{-1} + f_2 \alpha^{-2} + f_3 \alpha^{-3})
 + t_3 \alpha^{-3}$. $t_0$ and $t_3$ are found to be 9.39, -3.92. 
The FNV corrected formation 
energy shows excellent agreement with the extrapolated value for all the super cell 
sizes considered. Fig \ref{NR-fig2} (b) shows the dependence of the defect level 
accomodating the -1 charge with system size. The uncorrected values are fit with a 
polynomial of the form: 
$\epsilon_{-1}(\alpha) = t_0 + 2\times(f_1 \alpha^{-1} + f_2 \alpha^{-2} 
+ f_3 \alpha^{-3}) + t_3 \alpha^{-3}$. 
The corrected eigenvalues are in good agreement with the extrapolated value.

\section{Workflow}
\begin{figure}
 \centering
 \includegraphics[scale=0.3]{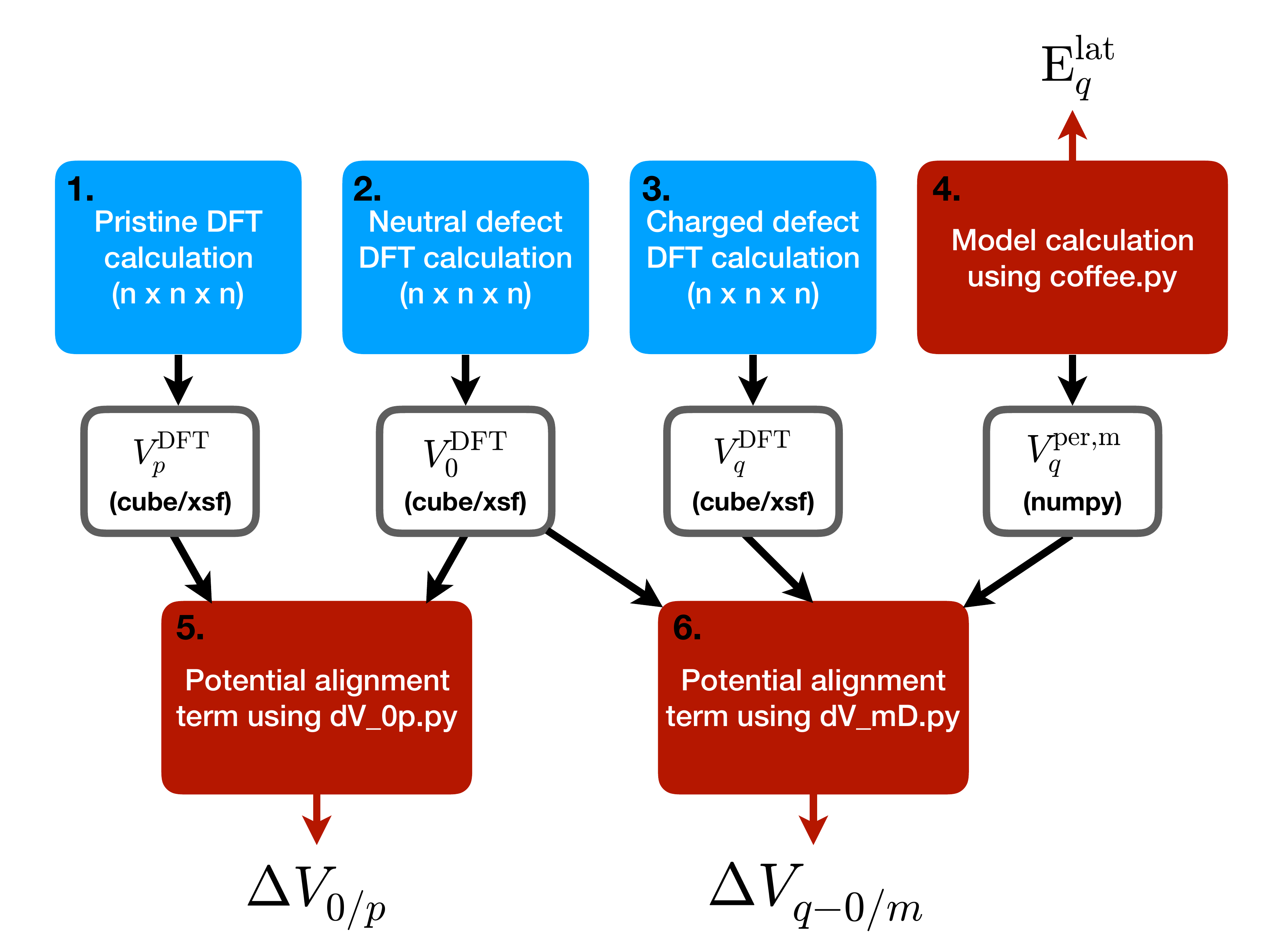}
 \caption {\label{workflow} (Color online)
  The workflow involved in computing the formation energy of 
  a charged defect. The blue boxed are performed using a 
  DFT electronic structure code. CoFFEE aids in computing 
  the corrections, the red boxes. (n$\times$n$\times$n) 
  refers to the super cell size.
  }
\end{figure}

The general
steps involved in computing corrections for the formation energy of
a charged defect are the following (Fig. \ref{workflow}):
\\

\begin{enumerate}
  \item Compute the total energy of the pristine super cell of the same size. Save the DFT potential
        in cube/xsf format.
  \item Compute the total energy of the super cell (say n$\times$n$\times$n) containing the
        neutral defect. Save the DFT potential in cube/xsf format.
  \item Compute the total energy of the super cell containing the
        charged defect. Save the DFT potential in cube/xsf format.
  \item Compute $\mathrm{E}^{\mathrm{lat}}_q$ term:
        Compute the model energy for various super cell sizes and
        extrapolate to obtain $\mathrm{E}^{\mathrm{iso,m}}_q$. $\mathrm{E}^{\mathrm{lat}}_q$
        is then given by: $\mathrm{E}^{\mathrm{iso,m}}_q - 
        \mathrm{E}^{\mathrm{per,m}}_q$ (n$\times$n$\times$n).
  \item Compute the potential alignment term $\Delta V_{0/p}$, Eqn. \ref{eqn_0p}.
        The utility script \emph{dV\_0p.py} can be used to compute this.
  \item Compute the potential alignment term $\Delta V_{q-0/m}$, Eqn. \ref{eqn5}.
        The utility \emph{dV\_mD.py} can be used to compute this.

\end{enumerate}

The first three steps are performed by the user with the DFT electronic
structure code of his/her choice.
The xsf/cube file formats are commonly used to visualize data. Most electronic structure
codes provide utilities to convert the DFT potential after a self consistent calculation
into these formats.
These formats act as an interface between the DFT calculation and the CoFFEE code.
The CoFFEE code aids in computing steps 4, 5 and 6. The model calculations are performed
by solving the Poisson equation, as detailed above. 
Details on preparing the input file and
running the CoFFEE code to perform these calculations are provided to some extent in 
the next section
and exhaustively in the user guide of the code. The user guide is available for download
with the distribution.

\section{CoFFEE code framework}

\subsection{Layout}
\begin{figure}
  \centering
  \includegraphics[scale=0.2]{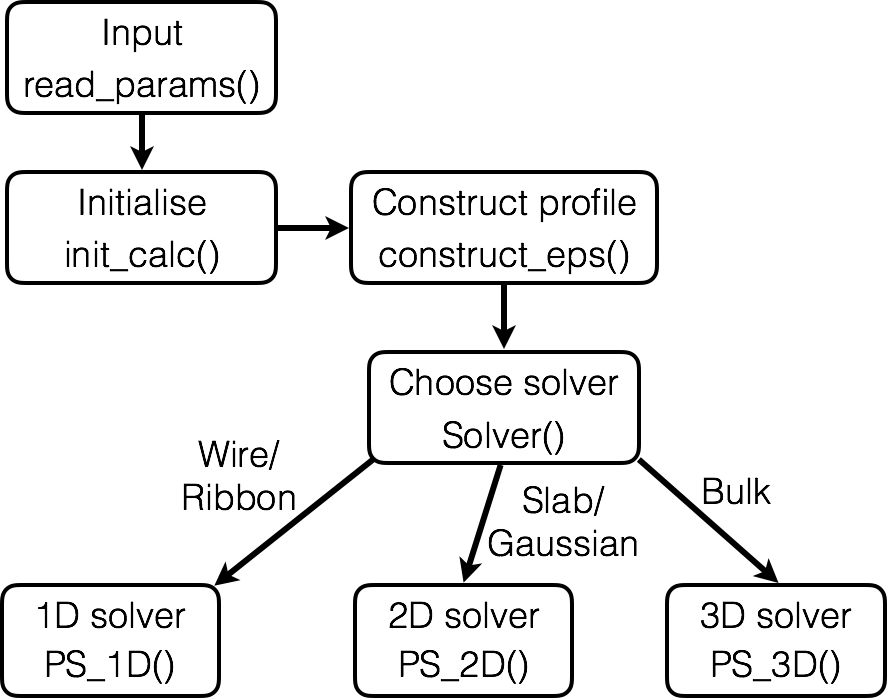}
  \caption {\label{Coffee_flow} 
  Program flow of the script \emph{coffee.py} to compute the total energy from a 
  model calculation using the CoFFEE code. 
           }
\end{figure}

On unzipping the tar file after download, the CoFFEE folder 
contains the following directories: 
\emph{PoissonSolver}, \emph{PotentialAlignment} and \emph{Examples},
and a script \emph{coffee.py}.
The script \emph{coffee.py} is the main executable. It can be called from
the user's working directory and it performs the 
model calculation by sequentially calling
the relevant functions as depicted in Fig \ref{Coffee_flow}. 
The script \emph{coffee.py} reads input from a file.
The input file contains parameters pertaining to the dimensions 
of the super cell, 
the dielectric profile and the model Gaussian charge. 
The next subsection describes the input file parameters.
On running the script, the model total energy is printed out.
The model potential,  dielectric profile and model charge distribution
are written as numpy save files on providing the appropriate flags 
in the input.
The script is run for different supercell sizes and the model energy extrapolated 
as shown in Fig. \ref{fig4} (c), \ref{MoS2-fig1} (c) and \ref{NR-fig1} 
(e) to obtain the isolated model energy.

The \emph{PoissonSolver} folder holds \emph{classes.py} which defines three classes: 
\emph{cell}, \emph{diel\_profile} and \emph{gaussian}. Each class has a 
function to read the parameters pertaining to it from the input file. 
The \emph{cell} class has information regarding the cell parameters and the plane wave 
cut off to be used in the calculation. The \emph{diel\_profile} class has attributes
regarding the dielectric profile to be used and functions to construct and
Fourier transform the requisite profile. The \emph{gaussian} class has attributes
regarding the construction and Fourier transform of the model gaussian charge.
The \emph{PoissonSolver} folder also consists of three routines, \emph{Solver.py}, 
\emph{construct\_eps.py} and \emph{PS\_main.py}.
\emph{construct\_eps.py} is used to select and construct the appropriate dielectric profile, 
based on the user input.
\emph{Solver.py} is used to select the poisson solver to be run. 
\emph{PS\_main.py} contains the solvers for 1D (\emph{PS\_1D}), 
2D (\emph{PS\_2D}) and bulk (\emph{PS\_3D}) systems.

\subsection{Input file}
The input file is divided into three sections designated with 
"\&CELL\_PARAMETERS", "\&DIELECTRIC\_PARAMETERS" and "\&GAUSSIAN\_PARAMETERS".
Each section is ended with a "/" and contains, in the intervening space, the
parameters relevant to the respective section. The following is an example of an input file 
for $V_{C}^{-2}$ in diamond.

\begin{verbatim}
&CELL_PARAMETERS
# Normalized lattice vectors: a1, a2 and a3
Lattice_Vectors(normalized):
1.000000000   0.000000000   0.000000000
0.000000000   1.000000000   0.000000000
0.000000000   0.000000000   1.000000000

# Cell dimensions. Provide "angstrom" in place 
# of "bohr" if you wish to specify
# these values in angstrom units.
# These are multiplied to a1, a2 and a3 respectively.
Cell_dimensions bohr
26.594331775231996 26.594331775231996 26.594331775231996

# G-vectors will be used upto this kinetic energy cut off. 
# Provide "Rydberg" in place of "Hartree"
# if you wish to specify the cut off in Rydberg 
# atomic units.
Ecut=20.0 Hartree
/

# Set "Bulk" here for 3D, bulk systems.
&DIELECTRIC_PARAMETERS Bulk
# Sets the value of the dielectric constant along a1, a2, a3. 
Epsilon1_a1 = 5.76
Epsilon1_a2 = 5.76
Epsilon1_a3 = 5.76
/

&GAUSSIAN_PARAMETERS:
# The charge state of the defect being simulated.
Total_charge = -2

# The width of the model Gaussian charge being used. 
# (default: bohr units)
Sigma = 2.614

# These set the center of the Gaussian in crystal 
# units.
Centre_a1 = 0.5
Centre_a2 = 0.5
Centre_a3 = 0.5
/
\end{verbatim}

The default unit of length is bohr, unless "angstrom" is explicitly specified after 
the number. Ecut stands for the plane wave energy cut off used in solving 
the Poisson equation.
The string after \&DIELECTRIC\_PARAMETERS determines the 
dielectric profile and which Poisson solver is to be called. The dielectric parameters 
are provided in the input file along the crystal axes. The center of the 
model charge is specified in crystal units in the \&GAUSSIAN\_PARAMETERS, along 
with the charge it carries and the gaussian width. For the nanowire profile, the 
user can provide a list of points that form the boundary of the cross-section
 of the wire in a file, based on which the profile is created. 
The user guide lists all the possible input 
parameters and the properties they control. Also, the \emph{Examples} folder has 
sample profiles for different materials. 

\subsection{Potential alignment}
The \emph{PotentialAlignment} directory holds two scripts: \emph{dV\_0p.py} and \emph{dV\_mD.py}
to compute the two potential alignment terms, $\Delta V_{0/p}$ and $\Delta V_{q-0/m}$ in Eqn.
\ref{eqn2} and \ref{eqn3}. These scripts
read the DFT potential from "cube" or "xsf" formats.
\emph{dV\_0p.py} plots the planar averaged $V_0 - V_p$, in the 
desired direction. The difference far from the defect can then be 
read from the plot. 
This script reads input from a file. The format of the 
input file:
\begin{verbatim}
&dV_0p
file_type = cube # No quotes. Takes cube/xsf
file_neutral = n.cube # No quotes. Path to the neutral DFT potential file
file_pristine = p.cube # No quotes. Path to the pristine DFT potential file
plt_dir = a1 # No quotes. Takes a1/a2/a3. If a1
             # is specified, the data is averaged along a2
             # and a3 directions and the planar averaged data is plotted
             # along a1 in a file pa_dv0p_a1.plot
factor = Ryd # factor to be multiplied to the cube/xsf data. If the data is in
             # rydberg and the plot is needed in eV, specify
             # factor = Ryd. If the data is in Hartree units,
             # specify factor = Hartree
/
\end{verbatim}

dV\_mD.py plots the planar averaged 
$(V_q^{\mathrm{DFT}} - V_0^{\mathrm{DFT}})$
and $V^{\mathrm{per,m}}_q |_{\mathrm{far}}$ 
 along the desired direction. These can be plotted as shown
in Fig. \ref{fig4} b, \ref{MoS2-fig1} b.
This script reads input from a file. The format of the
input file:

\begin{verbatim}
&dV_mD
file_type = cube # No quotes. Format of the DFT potential files: cube/xsf
file_model = m.npy # No quotes. Path to the model potential file (.npy)
file_charged = q.cube # No quotes. Path to the charged DFT potential file
file_neutral = n.cube # No quotes. Path to the neutral DFT potential file
plt_dir = a1 # No quotes. Takes a1/a2/a3. If a1
             # is specified, the data is averaged along a2
             # and a3 directions and the planar averaged data is plotted
             # along a1 in files DFTdiff_a1.plot, model_a1.plot
factor = Ryd # factor to be multiplied to the cube/xsf data. If the data is in
             # rydberg and the plot is needed in eV, specify
             # factor = Ryd. If the data is in Hartree units,
             # specify factor = Hartree
/
\end{verbatim}

\subsection{Solver parallelization and optimization}
\emph{PS\_1D}, \emph{PS\_2D} and \emph{PS\_3D} are the
poisson solver functions in \emph{PS\_main.py}
for 1D, 2D and bulk systems.
We do not use any parallelization for bulk systems, where the potential
is obtained using Eqn \ref{eqn8}, since the computation time is fairly small.
We have, however, optimized this function with the help of Cython \cite{cython}.
For 2D systems, linear equations of the form in Eqn \ref{eqn10} are solved.
The number of reciprocal lattice vectors depends on the plane wave energy cut off
set for the calculation. For $N_{G_1}$ and $N_{G_2}$ reciprocal lattice
vectors along the in-plane reciprocal lattice directions, a set of
$N_{G_1}\times N_{G_2}$ linear equations are solved. We use the iterative
solver \emph{bicgstab()} which is a part of the SciPy linear algebra package
to solve these linear equations.
These calculations are computationally intensive for large system sizes.
We parallelize the total number of linear equations so that each process solves
$N_{G_2}\times N_{G_1}/N_p$ equations, where $N_p$ is the number of processes.
The parallelization is done via MPI, using the package \emph{mpi4py} \cite{mpi4py}.
For 1D systems, $N_{G_z}$ linear equations are solved as shown in
Eqn \ref{eqn11} . These are again parallelized in the package in a similar manner.
Furthermore, the efficiency of the  \emph{bicgstab()} routine is primarily determined
 by the time taken to compute matrix vector products. We optimized these products using
Cython \cite{cython} to reduce the computation time.

\section{Conclusion}
We present a complete package, CoFFEE, for electrostatic corrections in charged 
defect simulations. We demonstrate the application of this code on three test systems, 
namely: bulk diamond, 2D MoS$_2$ and 1D BN nanoribbon. The corrected formation 
energy and defect eigenvalues for these systems are found to be in good agreement with 
the isolated limit extrapolation of the corresponding uncorrected quantities. The code, 
written completely in Python \cite{Rossum}, 
is parallelized using MPI and the slowest steps accelarated using Cython \cite{cython}. 

\section{Acknowledgments}
We thank the Supercomputer Education and Research Centre (SERC) at IISc 
for providing the computational facilities.







\section*{References}

\bibliographystyle{elsarticle-num}

\begin{thebibliography}{10}
\expandafter\ifx\csname url\endcsname\relax
  \def\url#1{\texttt{#1}}\fi
\expandafter\ifx\csname urlprefix\endcsname\relax\def\urlprefix{URL }\fi
\expandafter\ifx\csname href\endcsname\relax
  \def\href#1#2{#2} \def\path#1{#1}\fi

\bibitem{RMP_Walle}
C.~Freysoldt, B.~Grabowski, T.~Hickel, J.~Neugebauer, G.~Kresse, A.~Janotti,
  C.~G. Van~de Walle,
  \href{http://link.aps.org/doi/10.1103/RevModPhys.86.253}{First-principles
  calculations for point defects in solids}, Rev. Mod. Phys. 86 (2014)
  253--305.
\newblock \href {http://dx.doi.org/10.1103/RevModPhys.86.253}
  {\path{doi:10.1103/RevModPhys.86.253}}.
\newline\urlprefix\url{http://link.aps.org/doi/10.1103/RevModPhys.86.253}

\bibitem{PRB_Andrei}
A.~Malashevich, M.~Jain, S.~G. Louie,
  \href{http://link.aps.org/doi/10.1103/PhysRevB.89.075205}{First-principles
  $\mathrm{DFT}+gw$ study of oxygen vacancies in rutile ${\text{tio}}_{2}$},
  Phys. Rev. B 89 (2014) 075205.
\newblock \href {http://dx.doi.org/10.1103/PhysRevB.89.075205}
  {\path{doi:10.1103/PhysRevB.89.075205}}.
\newline\urlprefix\url{http://link.aps.org/doi/10.1103/PhysRevB.89.075205}

\bibitem{PRL_Jain}
M.~Jain, J.~R. Chelikowsky, S.~G. Louie,
  \href{http://link.aps.org/doi/10.1103/PhysRevLett.107.216803}{Quasiparticle
  excitations and charge transition levels of oxygen vacancies in hafnia},
  Phys. Rev. Lett. 107 (2011) 216803.
\newblock \href {http://dx.doi.org/10.1103/PhysRevLett.107.216803}
  {\path{doi:10.1103/PhysRevLett.107.216803}}.
\newline\urlprefix\url{http://link.aps.org/doi/10.1103/PhysRevLett.107.216803}

\bibitem{PRB_Choi}
S.~Choi, M.~Jain, S.~G. Louie,
  \href{http://link.aps.org/doi/10.1103/PhysRevB.86.041202}{Mechanism for
  optical initialization of spin in nv${}^{\ensuremath{-}}$ center in diamond},
  Phys. Rev. B 86 (2012) 041202.
\newblock \href {http://dx.doi.org/10.1103/PhysRevB.86.041202}
  {\path{doi:10.1103/PhysRevB.86.041202}}.
\newline\urlprefix\url{http://link.aps.org/doi/10.1103/PhysRevB.86.041202}

\bibitem{PRB_Bjaalie}
L.~Bjaalie, A.~Janotti, K.~Krishnaswamy, C.~G. Van~de Walle,
  \href{http://link.aps.org/doi/10.1103/PhysRevB.93.115316}{Point defects,
  impurities, and small hole polarons in ${\mathrm{gdtio}}_{3}$}, Phys. Rev. B
  93 (2016) 115316.
\newblock \href {http://dx.doi.org/10.1103/PhysRevB.93.115316}
  {\path{doi:10.1103/PhysRevB.93.115316}}.
\newline\urlprefix\url{http://link.aps.org/doi/10.1103/PhysRevB.93.115316}

\bibitem{PRA_Diallo}
I.~C. Diallo, D.~O. Demchenko,
  \href{http://link.aps.org/doi/10.1103/PhysRevApplied.6.064002}{Native point
  defects in gan: A hybrid-functional study}, Phys. Rev. Applied 6 (2016)
  064002.
\newblock \href {http://dx.doi.org/10.1103/PhysRevApplied.6.064002}
  {\path{doi:10.1103/PhysRevApplied.6.064002}}.
\newline\urlprefix\url{http://link.aps.org/doi/10.1103/PhysRevApplied.6.064002}

\bibitem{PRA_Oba}
Y.~Kumagai, L.~A. Burton, A.~Walsh, F.~Oba,
  \href{http://link.aps.org/doi/10.1103/PhysRevApplied.6.014009}{Electronic
  structure and defect physics of tin sulfides: Sns,
  ${\mathrm{sn}}_{2}{\mathrm{s}}_{3}$, and $\mathrm{Sn}{\mathrm{s}}_{2}$},
  Phys. Rev. Applied 6 (2016) 014009.
\newblock \href {http://dx.doi.org/10.1103/PhysRevApplied.6.014009}
  {\path{doi:10.1103/PhysRevApplied.6.014009}}.
\newline\urlprefix\url{http://link.aps.org/doi/10.1103/PhysRevApplied.6.014009}

\bibitem{PRB_Lee}
S.~R. Lee, A.~F. Wright, N.~A. Modine, C.~C. Battaile, S.~M. Foiles, J.~C.
  Thomas, A.~Van~der Ven,
  \href{http://link.aps.org/doi/10.1103/PhysRevB.92.045205}{First-principles
  survey of the structure, formation energies, and transition levels of
  as-interstitial defects in ingaas}, Phys. Rev. B 92 (2015) 045205.
\newblock \href {http://dx.doi.org/10.1103/PhysRevB.92.045205}
  {\path{doi:10.1103/PhysRevB.92.045205}}.
\newline\urlprefix\url{http://link.aps.org/doi/10.1103/PhysRevB.92.045205}

\bibitem{PRB_Wang}
V.~Wang, Y.~Kawazoe, W.~T. Geng,
  \href{http://link.aps.org/doi/10.1103/PhysRevB.91.045433}{Native point
  defects in few-layer phosphorene}, Phys. Rev. B 91 (2015) 045433.
\newblock \href {http://dx.doi.org/10.1103/PhysRevB.91.045433}
  {\path{doi:10.1103/PhysRevB.91.045433}}.
\newline\urlprefix\url{http://link.aps.org/doi/10.1103/PhysRevB.91.045433}

\bibitem{PRB_Sun}
W.~Sun, H.~Ehteshami, P.~A. Korzhavyi,
  \href{http://link.aps.org/doi/10.1103/PhysRevB.91.134111}{Structure and
  energy of point defects in tic: An \textit{ab initio} study}, Phys. Rev. B 91
  (2015) 134111.
\newblock \href {http://dx.doi.org/10.1103/PhysRevB.91.134111}
  {\path{doi:10.1103/PhysRevB.91.134111}}.
\newline\urlprefix\url{http://link.aps.org/doi/10.1103/PhysRevB.91.134111}

\bibitem{APL_Daniel}
D.~Steiauf, J.~L. Lyons, A.~Janotti, C.~G.~V. de~Walle,
  \href{http://dx.doi.org/10.1063/1.4894195}{First-principles study of
  vacancy-assisted impurity diffusion in zno}, APL Materials 2~(9) (2014)
  096101.
\newblock \href {http://arxiv.org/abs/http://dx.doi.org/10.1063/1.4894195}
  {\path{arXiv:http://dx.doi.org/10.1063/1.4894195}}, \href
  {http://dx.doi.org/10.1063/1.4894195} {\path{doi:10.1063/1.4894195}}.
\newline\urlprefix\url{http://dx.doi.org/10.1063/1.4894195}

\bibitem{JAP_Neugebauer}
C.~G.~V. de~Walle, J.~Neugebauer,
  \href{http://dx.doi.org/10.1063/1.1682673}{First-principles calculations for
  defects and impurities: Applications to iii-nitrides}, Journal of Applied
  Physics 95~(8) (2004) 3851--3879.
\newblock \href {http://arxiv.org/abs/http://dx.doi.org/10.1063/1.1682673}
  {\path{arXiv:http://dx.doi.org/10.1063/1.1682673}}, \href
  {http://dx.doi.org/10.1063/1.1682673} {\path{doi:10.1063/1.1682673}}.
\newline\urlprefix\url{http://dx.doi.org/10.1063/1.1682673}

\bibitem{PE_Mota}
F.~de~Brito~Mota, S.~Azevedo, C.~M. de~Castilho,
  \href{http://www.sciencedirect.com/science/article/pii/S1386947715301004}{Structural
  and electronic properties of perfect and defective bn nanoribbons: A dft
  study}, Physica E: Low-dimensional Systems and Nanostructures 74 (2015) 233
  -- 240.
\newblock \href
  {http://dx.doi.org/http://dx.doi.org/10.1016/j.physe.2015.06.028}
  {\path{doi:http://dx.doi.org/10.1016/j.physe.2015.06.028}}.
\newline\urlprefix\url{http://www.sciencedirect.com/science/article/pii/S1386947715301004}

\bibitem{JCTC_Chen}
W.~Chen, Y.~Li, G.~Yu, Z.~Zhou, Z.~Chen,
  \href{http://dx.doi.org/10.1021/ct900388x}{Electronic structure and
  reactivity of boron nitride nanoribbons with stone-wales defects}, Journal of
  Chemical Theory and Computation 5~(11) (2009) 3088--3095, pMID: 26609988.
\newblock \href {http://arxiv.org/abs/http://dx.doi.org/10.1021/ct900388x}
  {\path{arXiv:http://dx.doi.org/10.1021/ct900388x}}, \href
  {http://dx.doi.org/10.1021/ct900388x} {\path{doi:10.1021/ct900388x}}.
\newline\urlprefix\url{http://dx.doi.org/10.1021/ct900388x}

\bibitem{Manjanath.CPL}
A.~Manjanath, A.~K. Singh,
  \href{http://www.sciencedirect.com/science/article/pii/S0009261413014905}{Low
  formation energy and kinetic barrier of stone–wales defect in infinite and
  finite silicene}, Chemical Physics Letters 592 (2014) 52 -- 55.
\newblock \href
  {http://dx.doi.org/https://doi.org/10.1016/j.cplett.2013.12.010}
  {\path{doi:https://doi.org/10.1016/j.cplett.2013.12.010}}.
\newline\urlprefix\url{http://www.sciencedirect.com/science/article/pii/S0009261413014905}

\bibitem{PRL.Oif}
S.~\"O\ifmmode~\breve{g}\else \u{g}\fi{}\"ut, J.~R. Chelikowsky,
  \href{https://link.aps.org/doi/10.1103/PhysRevLett.91.235503}{Charge state
  dependent jahn-teller distortions of the $e$-center defect in crystalline
  si}, Phys. Rev. Lett. 91 (2003) 235503.
\newblock \href {http://dx.doi.org/10.1103/PhysRevLett.91.235503}
  {\path{doi:10.1103/PhysRevLett.91.235503}}.
\newline\urlprefix\url{https://link.aps.org/doi/10.1103/PhysRevLett.91.235503}

\bibitem{PRB.Tiago}
M.~L. Tiago, J.~R. Chelikowsky,
  \href{https://link.aps.org/doi/10.1103/PhysRevB.73.205334}{Optical
  excitations in organic molecules, clusters, and defects studied by
  first-principles green's function methods}, Phys. Rev. B 73 (2006) 205334.
\newblock \href {http://dx.doi.org/10.1103/PhysRevB.73.205334}
  {\path{doi:10.1103/PhysRevB.73.205334}}.
\newline\urlprefix\url{https://link.aps.org/doi/10.1103/PhysRevB.73.205334}

\bibitem{PRB.Canning}
J.~Kim, J.~W. Wilkins, F.~S. Khan, A.~Canning,
  \href{https://link.aps.org/doi/10.1103/PhysRevB.55.16186}{Extended si
  |p[311|p] defects}, Phys. Rev. B 55 (1997) 16186--16197.
\newblock \href {http://dx.doi.org/10.1103/PhysRevB.55.16186}
  {\path{doi:10.1103/PhysRevB.55.16186}}.
\newline\urlprefix\url{https://link.aps.org/doi/10.1103/PhysRevB.55.16186}

\bibitem{PRB_Komsa3}
H.-P. Komsa, S.~Kurasch, O.~Lehtinen, U.~Kaiser, A.~V. Krasheninnikov,
  \href{http://link.aps.org/doi/10.1103/PhysRevB.88.035301}{From point to
  extended defects in two-dimensional mos${}_{2}$: Evolution of atomic
  structure under electron irradiation}, Phys. Rev. B 88 (2013) 035301.
\newblock \href {http://dx.doi.org/10.1103/PhysRevB.88.035301}
  {\path{doi:10.1103/PhysRevB.88.035301}}.
\newline\urlprefix\url{http://link.aps.org/doi/10.1103/PhysRevB.88.035301}

\bibitem{JPCC_Talat}
D.~Le, T.~B. Rawal, T.~S. Rahman,
  \href{http://dx.doi.org/10.1021/jp411256g}{Single-layer mos2 with sulfur
  vacancies: Structure and catalytic application}, The Journal of Physical
  Chemistry C 118~(10) (2014) 5346--5351.
\newblock \href {http://arxiv.org/abs/http://dx.doi.org/10.1021/jp411256g}
  {\path{arXiv:http://dx.doi.org/10.1021/jp411256g}}, \href
  {http://dx.doi.org/10.1021/jp411256g} {\path{doi:10.1021/jp411256g}}.
\newline\urlprefix\url{http://dx.doi.org/10.1021/jp411256g}

\bibitem{ARMR_Tuller}
H.~L. Tuller, S.~R. Bishop,
  \href{http://dx.doi.org/10.1146/annurev-matsci-062910-100442}{Point defects
  in oxides: Tailoring materials through defect engineering}, Annual Review of
  Materials Research 41~(1) (2011) 369--398.
\newblock \href
  {http://arxiv.org/abs/http://dx.doi.org/10.1146/annurev-matsci-062910-100442}
  {\path{arXiv:http://dx.doi.org/10.1146/annurev-matsci-062910-100442}}, \href
  {http://dx.doi.org/10.1146/annurev-matsci-062910-100442}
  {\path{doi:10.1146/annurev-matsci-062910-100442}}.
\newline\urlprefix\url{http://dx.doi.org/10.1146/annurev-matsci-062910-100442}

\bibitem{AFM_Zhao}
L.~Hu, T.~Zhu, X.~Liu, X.~Zhao,
  \href{http://dx.doi.org/10.1002/adfm.201400474}{Point defect engineering of
  high-performance bismuth-telluride-based thermoelectric materials}, Advanced
  Functional Materials 24~(33) (2014) 5211--5218.
\newblock \href {http://dx.doi.org/10.1002/adfm.201400474}
  {\path{doi:10.1002/adfm.201400474}}.
\newline\urlprefix\url{http://dx.doi.org/10.1002/adfm.201400474}

\bibitem{JPCC_Nowotny}
M.~K. Nowotny, L.~R. Sheppard, T.~Bak, J.~Nowotny,
  \href{http://dx.doi.org/10.1021/jp077275m}{Defect chemistry of titanium
  dioxide. application of defect engineering in processing of tio2-based
  photocatalysts}, The Journal of Physical Chemistry C 112~(14) (2008)
  5275--5300.
\newblock \href {http://arxiv.org/abs/http://dx.doi.org/10.1021/jp077275m}
  {\path{arXiv:http://dx.doi.org/10.1021/jp077275m}}, \href
  {http://dx.doi.org/10.1021/jp077275m} {\path{doi:10.1021/jp077275m}}.
\newline\urlprefix\url{http://dx.doi.org/10.1021/jp077275m}

\bibitem{Singh.ACSNano}
A.~K. Singh, E.~S. Penev, B.~I. Yakobson,
  \href{http://dx.doi.org/10.1021/nn1006072}{Vacancy clusters in graphane as
  quantum dots}, ACS Nano 4~(6) (2010) 3510--3514.
\newblock \href {http://arxiv.org/abs/http://dx.doi.org/10.1021/nn1006072}
  {\path{arXiv:http://dx.doi.org/10.1021/nn1006072}}, \href
  {http://dx.doi.org/10.1021/nn1006072} {\path{doi:10.1021/nn1006072}}.
\newline\urlprefix\url{http://dx.doi.org/10.1021/nn1006072}

\bibitem{PRB.Bhowmick}
S.~Bhowmick, U.~V. Waghmare,
  \href{https://link.aps.org/doi/10.1103/PhysRevB.81.155416}{Anisotropy of the
  stone-wales defect and warping of graphene nanoribbons: A first-principles
  analysis}, Phys. Rev. B 81 (2010) 155416.
\newblock \href {http://dx.doi.org/10.1103/PhysRevB.81.155416}
  {\path{doi:10.1103/PhysRevB.81.155416}}.
\newline\urlprefix\url{https://link.aps.org/doi/10.1103/PhysRevB.81.155416}

\bibitem{PRL_FNV}
C.~Freysoldt, J.~Neugebauer, C.~G. Van~de Walle,
  \href{http://link.aps.org/doi/10.1103/PhysRevLett.102.016402}{Fully
  \textit{Ab Initio} finite-size corrections for charged-defect supercell
  calculations}, Phys. Rev. Lett. 102 (2009) 016402.
\newblock \href {http://dx.doi.org/10.1103/PhysRevLett.102.016402}
  {\path{doi:10.1103/PhysRevLett.102.016402}}.
\newline\urlprefix\url{http://link.aps.org/doi/10.1103/PhysRevLett.102.016402}

\bibitem{PRB_KRP}
H.-P. Komsa, T.~T. Rantala, A.~Pasquarello,
  \href{http://link.aps.org/doi/10.1103/PhysRevB.86.045112}{Finite-size
  supercell correction schemes for charged defect calculations}, Phys. Rev. B
  86 (2012) 045112.
\newblock \href {http://dx.doi.org/10.1103/PhysRevB.86.045112}
  {\path{doi:10.1103/PhysRevB.86.045112}}.
\newline\urlprefix\url{http://link.aps.org/doi/10.1103/PhysRevB.86.045112}

\bibitem{PRB_Oba}
Y.~Kumagai, F.~Oba,
  \href{http://link.aps.org/doi/10.1103/PhysRevB.89.195205}{Electrostatics-based
  finite-size corrections for first-principles point defect calculations},
  Phys. Rev. B 89 (2014) 195205.
\newblock \href {http://dx.doi.org/10.1103/PhysRevB.89.195205}
  {\path{doi:10.1103/PhysRevB.89.195205}}.
\newline\urlprefix\url{http://link.aps.org/doi/10.1103/PhysRevB.89.195205}

\bibitem{PRB_Dabo}
I.~Dabo, B.~Kozinsky, N.~E. Singh-Miller, N.~Marzari,
  \href{http://link.aps.org/doi/10.1103/PhysRevB.77.115139}{Electrostatics in
  periodic boundary conditions and real-space corrections}, Phys. Rev. B 77
  (2008) 115139.
\newblock \href {http://dx.doi.org/10.1103/PhysRevB.77.115139}
  {\path{doi:10.1103/PhysRevB.77.115139}}.
\newline\urlprefix\url{http://link.aps.org/doi/10.1103/PhysRevB.77.115139}

\bibitem{JPCSSP_Gillian}
M.~Leslie, N.~J. Gillan,
  \href{http://stacks.iop.org/0022-3719/18/i=5/a=005}{The energy and elastic
  dipole tensor of defects in ionic crystals calculated by the supercell
  method}, Journal of Physics C: Solid State Physics 18~(5) (1985) 973.
\newline\urlprefix\url{http://stacks.iop.org/0022-3719/18/i=5/a=005}

\bibitem{PRB_Hine}
N.~D.~M. Hine, K.~Frensch, W.~M.~C. Foulkes, M.~W. Finnis,
  \href{http://link.aps.org/doi/10.1103/PhysRevB.79.024112}{Supercell size
  scaling of density functional theory formation energies of charged defects},
  Phys. Rev. B 79 (2009) 024112.
\newblock \href {http://dx.doi.org/10.1103/PhysRevB.79.024112}
  {\path{doi:10.1103/PhysRevB.79.024112}}.
\newline\urlprefix\url{http://link.aps.org/doi/10.1103/PhysRevB.79.024112}

\bibitem{JPCM_Chen}
W.~Chen, A.~Pasquarello,
  \href{http://stacks.iop.org/0953-8984/27/i=13/a=133202}{First-principles
  determination of defect energy levels through hybrid density functionals and
  gw}, Journal of Physics: Condensed Matter 27~(13) (2015) 133202.
\newline\urlprefix\url{http://stacks.iop.org/0953-8984/27/i=13/a=133202}

\bibitem{PRB_CP}
W.~Chen, A.~Pasquarello,
  \href{http://link.aps.org/doi/10.1103/PhysRevB.88.115104}{Correspondence of
  defect energy levels in hybrid density functional theory and many-body
  perturbation theory}, Phys. Rev. B 88 (2013) 115104.
\newblock \href {http://dx.doi.org/10.1103/PhysRevB.88.115104}
  {\path{doi:10.1103/PhysRevB.88.115104}}.
\newline\urlprefix\url{http://link.aps.org/doi/10.1103/PhysRevB.88.115104}

\bibitem{PRB_Payne}
G.~Makov, M.~C. Payne,
  \href{http://link.aps.org/doi/10.1103/PhysRevB.51.4014}{Periodic boundary
  conditions in \textit{ab initio} calculations}, Phys. Rev. B 51 (1995)
  4014--4022.
\newblock \href {http://dx.doi.org/10.1103/PhysRevB.51.4014}
  {\path{doi:10.1103/PhysRevB.51.4014}}.
\newline\urlprefix\url{http://link.aps.org/doi/10.1103/PhysRevB.51.4014}

\bibitem{PRB_Zunger}
S.~Lany, A.~Zunger,
  \href{http://link.aps.org/doi/10.1103/PhysRevB.78.235104}{Assessment of
  correction methods for the band-gap problem and for finite-size effects in
  supercell defect calculations: Case studies for zno and gaas}, Phys. Rev. B
  78 (2008) 235104.
\newblock \href {http://dx.doi.org/10.1103/PhysRevB.78.235104}
  {\path{doi:10.1103/PhysRevB.78.235104}}.
\newline\urlprefix\url{http://link.aps.org/doi/10.1103/PhysRevB.78.235104}

\bibitem{PRL_Zhang}
D.~Wang, D.~Han, X.-B. Li, S.-Y. Xie, N.-K. Chen, W.~Q. Tian, D.~West, H.-B.
  Sun, S.~B. Zhang,
  \href{http://link.aps.org/doi/10.1103/PhysRevLett.114.196801}{Determination
  of formation and ionization energies of charged defects in two-dimensional
  materials}, Phys. Rev. Lett. 114 (2015) 196801.
\newblock \href {http://dx.doi.org/10.1103/PhysRevLett.114.196801}
  {\path{doi:10.1103/PhysRevLett.114.196801}}.
\newline\urlprefix\url{http://link.aps.org/doi/10.1103/PhysRevLett.114.196801}

\bibitem{CPC.Chan}
T.-L. Chan, A.~J. Lee, J.~R. Chelikowsky,
  \href{http://www.sciencedirect.com/science/article/pii/S0010465514000642}{An
  effective capacitance model for computing the electronic properties of
  charged defects in crystals}, Computer Physics Communications 185~(6) (2014)
  1564 -- 1569.
\newblock \href {http://dx.doi.org/https://doi.org/10.1016/j.cpc.2014.02.020}
  {\path{doi:https://doi.org/10.1016/j.cpc.2014.02.020}}.
\newline\urlprefix\url{http://www.sciencedirect.com/science/article/pii/S0010465514000642}

\bibitem{PRL_KP}
H.-P. Komsa, A.~Pasquarello,
  \href{http://link.aps.org/doi/10.1103/PhysRevLett.110.095505}{Finite-size
  supercell correction for charged defects at surfaces and interfaces}, Phys.
  Rev. Lett. 110 (2013) 095505.
\newblock \href {http://dx.doi.org/10.1103/PhysRevLett.110.095505}
  {\path{doi:10.1103/PhysRevLett.110.095505}}.
\newline\urlprefix\url{http://link.aps.org/doi/10.1103/PhysRevLett.110.095505}

\bibitem{PRB_Noh}
J.-Y. Noh, H.~Kim, Y.-S. Kim,
  \href{http://link.aps.org/doi/10.1103/PhysRevB.89.205417}{Stability and
  electronic structures of native defects in single-layer
  ${\mathrm{mos}}_{2}$}, Phys. Rev. B 89 (2014) 205417.
\newblock \href {http://dx.doi.org/10.1103/PhysRevB.89.205417}
  {\path{doi:10.1103/PhysRevB.89.205417}}.
\newline\urlprefix\url{http://link.aps.org/doi/10.1103/PhysRevB.89.205417}

\bibitem{PRX_Komsa}
H.-P. Komsa, N.~Berseneva, A.~V. Krasheninnikov, R.~M. Nieminen,
  \href{http://link.aps.org/doi/10.1103/PhysRevX.4.031044}{Charged point
  defects in the flatland: Accurate formation energy calculations in
  two-dimensional materials}, Phys. Rev. X 4 (2014) 031044.
\newblock \href {http://dx.doi.org/10.1103/PhysRevX.4.031044}
  {\path{doi:10.1103/PhysRevX.4.031044}}.
\newline\urlprefix\url{http://link.aps.org/doi/10.1103/PhysRevX.4.031044}

\bibitem{PRB_Komsa2}
H.-P. Komsa, A.~V. Krasheninnikov,
  \href{http://link.aps.org/doi/10.1103/PhysRevB.91.125304}{Native defects in
  bulk and monolayer ${\mathrm{mos}}_{2}$ from first principles}, Phys. Rev. B
  91 (2015) 125304.
\newblock \href {http://dx.doi.org/10.1103/PhysRevB.91.125304}
  {\path{doi:10.1103/PhysRevB.91.125304}}.
\newline\urlprefix\url{http://link.aps.org/doi/10.1103/PhysRevB.91.125304}

\bibitem{PRB_Park}
S.~Kim, K.~J. Chang, J.-S. Park,
  \href{http://link.aps.org/doi/10.1103/PhysRevB.90.085435}{Finite-size
  supercell correction scheme for charged defects in one-dimensional systems},
  Phys. Rev. B 90 (2014) 085435.
\newblock \href {http://dx.doi.org/10.1103/PhysRevB.90.085435}
  {\path{doi:10.1103/PhysRevB.90.085435}}.
\newline\urlprefix\url{http://link.aps.org/doi/10.1103/PhysRevB.90.085435}

\bibitem{arxiv.Kax}
D.~{Vinichenko}, M.~{Gokhan Sensoy}, C.~M. {Friend}, E.~{Kaxiras}, {Accurate
  formation energies of charged defects in solids: a systematic approach},
  ArXiv e-prints\href {http://arxiv.org/abs/1701.02521}
  {\path{arXiv:1701.02521}}.

\bibitem{PRB.Sensoy}
M.~G. Sensoy, D.~Vinichenko, W.~Chen, C.~M. Friend, E.~Kaxiras,
  \href{https://link.aps.org/doi/10.1103/PhysRevB.95.014106}{Strain effects on
  the behavior of isolated and paired sulfur vacancy defects in monolayer
  ${\mathrm{mos}}_{2}$}, Phys. Rev. B 95 (2017) 014106.
\newblock \href {http://dx.doi.org/10.1103/PhysRevB.95.014106}
  {\path{doi:10.1103/PhysRevB.95.014106}}.
\newline\urlprefix\url{https://link.aps.org/doi/10.1103/PhysRevB.95.014106}

\bibitem{arxiv.Bro}
D.~{Broberg}, B.~{Medasani}, N.~{Zimmermann}, A.~{Canning}, M.~{Haranczyk},
  M.~{Asta}, G.~{Hautier}, {PyCDT: A Python toolkit for modeling point defects
  in semiconductors and insulators}, ArXiv e-prints\href
  {http://arxiv.org/abs/1611.07481} {\path{arXiv:1611.07481}}.

\bibitem{CPL.PyDEF}
E.~Péan, J.~Vidal, S.~Jobic, C.~Latouche,
  \href{http://www.sciencedirect.com/science/article/pii/S0009261417300015}{Presentation
  of the pydef post-treatment python software to compute publishable charts for
  defect energy formation}, Chemical Physics Letters 671~(Supplement C) (2017)
  124 -- 130.
\newblock \href
  {http://dx.doi.org/https://doi.org/10.1016/j.cplett.2017.01.001}
  {\path{doi:https://doi.org/10.1016/j.cplett.2017.01.001}}.
\newline\urlprefix\url{http://www.sciencedirect.com/science/article/pii/S0009261417300015}

\bibitem{Rossum}
G.~Rossum, Python reference manual, Tech. rep., Amsterdam, The Netherlands, The
  Netherlands (1995).

\bibitem{cython}
S.~Behnel, R.~Bradshaw, C.~Citro, L.~Dalcin, D.~Seljebotn, K.~Smith, Cython:
  The best of both worlds, Computing in Science Engineering 13~(2) (2011) 31
  --39.
\newblock \href {http://dx.doi.org/10.1109/MCSE.2010.118}
  {\path{doi:10.1109/MCSE.2010.118}}.

\bibitem{PRB_Gonze}
X.~Gonze, C.~Lee,
  \href{http://link.aps.org/doi/10.1103/PhysRevB.55.10355}{Dynamical matrices,
  born effective charges, dielectric permittivity tensors, and interatomic
  force constants from density-functional perturbation theory}, Phys. Rev. B 55
  (1997) 10355--10368.
\newblock \href {http://dx.doi.org/10.1103/PhysRevB.55.10355}
  {\path{doi:10.1103/PhysRevB.55.10355}}.
\newline\urlprefix\url{http://link.aps.org/doi/10.1103/PhysRevB.55.10355}

\bibitem{RMP_Baroni}
S.~Baroni, S.~de~Gironcoli, A.~Dal~Corso, P.~Giannozzi,
  \href{http://link.aps.org/doi/10.1103/RevModPhys.73.515}{Phonons and related
  crystal properties from density-functional perturbation theory}, Rev. Mod.
  Phys. 73 (2001) 515--562.
\newblock \href {http://dx.doi.org/10.1103/RevModPhys.73.515}
  {\path{doi:10.1103/RevModPhys.73.515}}.
\newline\urlprefix\url{http://link.aps.org/doi/10.1103/RevModPhys.73.515}

\bibitem{QE.Giannozi}
P.~Giannozzi, S.~Baroni, N.~Bonini, M.~Calandra, R.~Car, C.~Cavazzoni,
  D.~Ceresoli, G.~L. Chiarotti, M.~Cococcioni, I.~Dabo, A.~D. Corso,
  S.~de~Gironcoli, S.~Fabris, G.~Fratesi, R.~Gebauer, U.~Gerstmann,
  C.~Gougoussis, A.~Kokalj, M.~Lazzeri, L.~Martin-Samos, N.~Marzari, F.~Mauri,
  R.~Mazzarello, S.~Paolini, A.~Pasquarello, L.~Paulatto, C.~Sbraccia,
  S.~Scandolo, G.~Sclauzero, A.~P. Seitsonen, A.~Smogunov, P.~Umari, R.~M.
  Wentzcovitch, Quantum espresso: a modular and open-source software project
  for quantum simulations of materials, Journal of Physics: Condensed Matter
  21~(39) (2009) 395502.

\bibitem{PRB.Perdew}
J.~P. Perdew, A.~tunger, Self-interaction correction to density-functional
  approximations for many-electron systems, Phys. Rev. B 23 (1981) 5048--5079.
\newblock \href {http://dx.doi.org/10.1103/PhysRevB.23.5048}
  {\path{doi:10.1103/PhysRevB.23.5048}}.

\bibitem{PRB.PAW}
P.~E. Bl\"ochl,
  \href{http://link.aps.org/doi/10.1103/PhysRevB.50.17953}{Projector
  augmented-wave method}, Phys. Rev. B 50 (1994) 17953--17979.
\newblock \href {http://dx.doi.org/10.1103/PhysRevB.50.17953}
  {\path{doi:10.1103/PhysRevB.50.17953}}.
\newline\urlprefix\url{http://link.aps.org/doi/10.1103/PhysRevB.50.17953}

\bibitem{JCP.PBE}
J.~P. Perdew, M.~Ernzerhof, K.~Burke, Rationale for mixing exact exchange with
  density functional approximations, The Journal of Chemical Physics 105~(22)
  (1996) 9982--9985.
\newblock \href {http://dx.doi.org/http://dx.doi.org/10.1063/1.472933}
  {\path{doi:http://dx.doi.org/10.1063/1.472933}}.

\bibitem{mpi4py}
L.~Dalc\'{\i}n, R.~Paz, M.~Storti,
  \href{http://dx.doi.org/10.1016/j.jpdc.2005.03.010}{Mpi for python}, J.
  Parallel Distrib. Comput. 65~(9) (2005) 1108--1115.
\newblock \href {http://dx.doi.org/10.1016/j.jpdc.2005.03.010}
  {\path{doi:10.1016/j.jpdc.2005.03.010}}.
\newline\urlprefix\url{http://dx.doi.org/10.1016/j.jpdc.2005.03.010}

\end{thebibliography}







\end{document}